\documentclass[a4paper,fleqn,usenatbib]{mnras}

\usepackage{newtxtext,newtxmath}
\usepackage[T1]{fontenc}
\usepackage{ae,aecompl}
\usepackage{graphicx}
\usepackage{amsmath}
\usepackage{adjustbox}

\newcommand{\aref}[1]{\hyperref[#1]{Appendix~\ref{#1}}}

\newcommand*\samethanks[1][\value{footnote}]{\footnotemark[#1]}

\newcommand{\Msun}{\mbox{$M_{\odot}$}}

\newcommand{\HST}{\mbox{\textit{HST}}}
\newcommand{\EBV}{\mbox{$E(B-V)$}}
\newcommand{\EBVg}{\mbox{$E(B-V)_{\text{gas}}$}}
\newcommand{\EBVs}{\mbox{$E(B-V)_{\text{stars}}$}}
\newcommand{\EBVgtot}{\mbox{$E(B-V)_{\text{gas,global}}$}}
\newcommand{\EBVstot}{\mbox{$E(B-V)_{\text{stars,global}}$}}
\newcommand{\EBVgslit}{\mbox{$E(B-V)_{\text{gas,slit}}$}}
\newcommand{\EBVsslit}{\mbox{$E(B-V)_{\text{stars,slit}}$}}
\newcommand{\SFRha}{\mbox{SFR$_{\text{H}\alpha}$}}
\newcommand{\SFRhatot}{\mbox{SFR$_{\text{H}\alpha\text{,global}}$}}
\newcommand{\SFRhaslit}{\mbox{SFR$_{\text{H}\alpha\text{,slit}}$}}
\newcommand{\SFRuv}{\mbox{SFR$_{\text{UV}}$}}
\newcommand{\SFRuvtot}{\mbox{SFR$_{\text{UV,global}}$}}
\newcommand{\SFRuvslit}{\mbox{SFR$_{\text{UV,slit}}$}}
\newcommand{\SFRslit}{\mbox{SFR$_{\text{slit}}$}}
\newcommand{\SFRtot}{\mbox{SFR$_{\text{global}}$}}
\newcommand{\slitfrac}{\mbox{A$_{\text{slit}}$/A$_{\text{global}}$}}
\newcommand{\HA}{\mbox{H$\alpha$}}
\newcommand{\HB}{\mbox{H$\beta$}}

\newcommand{\insamp}{675}
\newcommand{\pnsamp}{449}
\newcommand{\nsamp}{303}
\newcommand{\izmin}{1.24}
\newcommand{\izmax}{2.66}
\newcommand{\zmin}{1.36}
\newcommand{\zmax}{2.66}

\title[SFR Ratios at $z\sim2$]{The MOSDEF Survey: The Dependence of \HA-to-UV SFR Ratios on SFR and Size at $z\sim2$\thanks{Some of the data presented herein were obtained at the W. M. Keck Observatory, which is operated as a scientific partnership among the California Institute of Technology, the University of California, and the National Aeronautics and Space Administration. The Observatory was made possible by the generous financial support of the W. M. Keck Foundation.}}

\author[T. Fetherolf et al.]{Tara Fetherolf,$^1$\thanks{E-mail: Tara.Fetherolf@gmail.com (TF)}
Naveen A. Reddy,$^1$
Alice E. Shapley,$^2$
Mariska Kriek,$^3$
\newauthor
Brian Siana,$^1$
Alison L. Coil,$^4$
Bahram Mobasher,$^1$
William R. Freeman,$^1$
\newauthor
Ryan L. Sanders,$^5$\thanks{Hubble Fellow}
Sedona H. Price,$^6$
Irene Shivaei,$^7$\samethanks\
Mojegan Azadi,$^8$
\newauthor
Laura de Groot,$^9$
Gene C.K. Leung,$^{10}$
and Tom O. Zick$^3$
\\
$^1$Department of Physics \& Astronomy, University of California, Riverside, 900 University Ave., Riverside, CA 92521, USA \\
$^2$Department of Physics \& Astronomy, University of California, Los Angeles, 430 Portola Plaza, Los Angeles, CA 90095, USA \\
$^3$Astronomy Department, University of California at Berkeley, Berkeley, CA 94720, USA \\
$^4$Center for Astrophysics and Space Sciences, Department of Physics, University of California, San Diego, 9500 Gilman Drive, La Jolla, CA 92093, USA \\
$^5$Department of Physics, University of California, Davis, 1 Shields Avenue, Davis, CA 95616, USA \\
$^6$Max-Planck-Institut F\"{u}r Extraterrestrische Physik, Postfach 1312, Garching, D-85741, Germany \\
$^7$Steward Observatory, University of Arizona, 933 North Cherry Avenue, Tucson, AZ 85721, USA \\
$^8$Center for Astrophysics | Harvard \& Smithsonian, 60 Garden Street, Cambridge, MA 02138, USA \\
$^9$Department of Physics, The College of Wooster, 1189 Beall Avenue, Wooster, OH 44691, USA \\
$^{10}$Department of Astronomy, University of Texas at Austin, Austin, TX, 78712
}

\pubyear{2021}

\begin{document}
\label{firstpage}
\pagerange{\pageref{firstpage}--\pageref{lastpage}}
\maketitle

\begin{abstract}
%
We perform an aperture-matched analysis of dust-corrected \HA\ and UV SFRs using \nsamp\ star-forming galaxies with spectroscopic redshifts $\zmin<z_\text{spec}<\zmax$ from the MOSFIRE Deep Evolution Field (MOSDEF) survey. By combining \HA\ and \HB\ emission line measurements with multi-waveband resolved CANDELS/3D-HST imaging, we directly compare dust-corrected \HA\ and UV SFRs, inferred assuming a fixed attenuation curve shape and constant SFHs, within the spectroscopic aperture. Previous studies have found that \HA\ and UV SFRs inferred with these assumptions generally agree for typical star-forming galaxies, but become increasingly discrepant for galaxies with higher SFRs ($\gtrsim$100\,\Msun\,yr$^{-1}$), with \HA-to-UV SFR ratios being larger for these galaxies. Our analysis shows that this trend persists even after carefully accounting for the apertures over which \HA\ and UV-based SFRs (and the nebular and stellar continuum reddening) are derived. Furthermore, our results imply that \HA\ SFRs may be higher in the centers of large galaxies (i.e., where there is coverage by the spectroscopic aperture) compared to their outskirts, which could be indicative of inside-out galaxy growth. Overall, we suggest that the persistent difference between nebular and stellar continuum reddening and high \HA-to-UV SFR ratios at the centers of large galaxies may be indicative of a patchier distribution of dust in galaxies with high SFRs. 
\end{abstract}

\section{Introduction} 
Galaxies were rapidly assembling their stellar mass at $z\sim2$ when cosmic star-formation activity was at its peak \citep[see][]{Madau14}. An important measure of how galaxies build their stellar mass is their star-formation rate (SFR). The SFR of galaxies in the local universe has been probed by observations of the ultraviolet (UV) continuum, nebular emission lines (e.g., \HA, \HB, Br$\gamma$), mid- and far-infrared (IR) continuum, radio continuum, X-ray emission, and combinations of these tracers \citep[e.g.,][]{Kennicutt98-1, Kennicutt12, Hao11}.  SFR measurements for higher redshift galaxies, however, typically rely on a limited set of tracers.  For example, at higher redshifts, the longer-wavelength tracers are of limited use as typical unlensed galaxies are not directly detected at these wavelengths. As such, we must rely on SFR indicators that can be accessed for individual high-redshift galaxies (e.g., UV continuum, \HA, \HB), but which may require corrections for the obscuring effects of dust. 

Dust obscuration can be directly probed by comparing the measured Balmer decrement (\HA/\HB) to the intrinsic Balmer decrement expected from typical conditions in the interstellar medium \citep[ISM;][]{Osterbrock89}.  This method is sensitive to the reddening towards the youngest stellar populations that host the most massive stars. An alternative method for estimating dust obscuration is to measure the slope of the UV continuum \citep{Calzetti94, Meurer99}, which is sensitive to stars over a broader range of stellar mass compared to the Balmer decrement. In general, these two measures for obscuration cannot be used interchangeably. Several studies of high-redshift galaxies have noted higher reddening towards the most massive stars (\EBVg) compared to the stellar continuum (\EBVs) by a constant factor \citep[e.g.,][]{Meurer99, Calzetti00, Kashino13, Price14, Tacchella18} or as a function of SFR \citep{Yoshikawa10, Kreckel13, Reddy15}, specific SFR \citep{Wild11, Price14, Boquien15, Reddy15, Hemmati15}, SFR surface density \citep{Boquien15}, and gas-phase metallicity \citep{Shivaei20}. 

Once appropriate dust corrections have been applied, \HA\ and UV SFRs have been found to generally agree for most typical star-forming galaxies at high redshifts \citep[e.g.,][]{Erb06-1, Reddy10, Wuyts13, Steidel14, Reddy15, Shivaei15-1, Shivaei20, Theios19, Shivaei20}. However, some authors have found that SFRs derived from nebular emission lines are systematically higher than those inferred from the UV continuum in high-redshift galaxies with high SFRs \citep[$\gtrsim$100\,\Msun\,yr$^{-1}$;][]{Hao11, Reddy15, Katsianis17}. Variations in SFRs derived using different tracers may result from their sensitivity to star formation on different timescales \citep{Reddy12-1, Price14}, which could provide useful information about the initial mass function (IMF) or star-formation history (SFH) of galaxies \citep{Madau14}, or could alternatively be caused by incorrect assumptions regarding dust corrections.

A non-uniform distribution of dust and stars has been invoked to explain the observed differences between nebular and stellar continuum reddening \citep{Calzetti94, Wild11, Price14, Reddy20}, and could also potentially explain the observed differences between SFRs inferred from \HA\ and UV data \citep{Boquien09, Boquien15, Hao11, Reddy15, Katsianis17}. The young massive stars that dominate the nebular line emission \citep[$\gtrsim$15 \Msun;][]{Kennicutt12} may be enshrouded in their parent birth clouds, while the UV continuum is dominated by stars over a larger range of main sequence lifetimes \citep[10--200 Myr;][]{Kennicutt12}. However, some of the less massive O and B stars will survive the dissipation of their birth clouds \citep{Calzetti94, Boquien15, Faisst17, Narayanan18} and will disproportionally add to the UV continuum light compared to more obscured OB associations. Furthermore, the nebular emission coming from the OB associations that have escaped their heavily obscured birth clouds will be overall less dust reddened---with reddening that is more similar to that which is typical of the UV continuum---compared to younger OB associations. Therefore, the dust that is reddening the nebular emission could also originate from the ionized or neutral gas phase of the ISM, such that the differences between the nebular and stellar continuum reddening may not necessarily be exclusively connected to the dust proximate to the youngest OB associations \citep{Reddy20}.

In addition to causing inconsistent SFR and reddening estimates, if geometrical effects within galaxies are at play, then the assumptions that are used to correct spectroscopic emission line measurements for light falling outside of the slit aperture may not be robust \citep{Brinchmann04, Kewley05, Salim07, Richards16, Green17}. Putting aside the dust corrections, if the spectroscopic emission line measurements are not corrected for slit-loss, then \HA\ SFRs, which are typically determined from spectroscopic observations, would lie systematically lower than UV SFRs, which are typically inferred from broadband photometry. Additionally, total \HA\ SFRs could also be over or underestimated if inappropriate assumptions are made about the Balmer decrement or if one assumes a flat \HA\ light profile, in contrast to the centrally-peaked \HA\ profiles observed in $z\sim2$ galaxies where \HA\ surface brightness decreases with increasing galactocentric distance \citep[e.g.,][]{Nelson13, Nelson16, Hemmati15, Tacchella18}. Furthermore, the average Balmer decrement measured across the spectroscopic slit may not necessarily be indicative of the globally-averaged Balmer decrement and could contribute to any discrepancies between \EBVg\ and the globally-measured \EBVs. Therefore, it is important to identify the significance of any ``aperture'' biases when comparing spectroscopic measurements of nebular emission lines with measurements of the UV continuum, the latter of which is typically performed on the whole galaxy.

In this study, we investigate how aperture effects influence the interpretation of reddening and SFRs derived from the UV continuum and nebular emission line observations. We pair rest-frame optical spectra from the MOSFIRE Deep Evolution Field survey \citep[MOSDEF;][]{Kriek15} with multi-waveband resolved imaging from the Cosmic Assembly Near-infrared Deep Extragalactic Legacy Survey \citep[CANDELS;][]{Grogin11, Koekemoer11} and the broadband photometric catalog compiled by the 3D-HST survey \citep{Brammer12, Skelton14}. The MOSDEF survey obtained rest-frame optical spectra for $\sim$1500~star-forming and AGN galaxies at $1.4\lesssim z \lesssim3.8$. Here we take advantage of the statistically large MOSDEF spectroscopic survey by analyzing a sample of \nsamp\ star-forming galaxies with detected \HA\ and \HB\ emission at $z\sim2$. 

The data and sample selection are presented in \autoref{sec:data}. In \autoref{sec:methods}, we describe our methodology for measuring dust-corrected \HA\ and UV SFRs directly within the MOSFIRE spectroscopic slit region. We then compare the globally measured dust-corrected SFRs to those measured inside the spectroscopic aperture in \autoref{sec:slit} and discuss our findings in \autoref{sec:discussion}. Finally, our results are summarized in \autoref{sec:summary}.

Throughout this paper, we assume a cosmology with $H_0=70$\,km\,s$^{-1}$\,Mpc$^{-1}$, ${\Omega_{\Lambda}=0.7}$, and ${\Omega_m=0.3}$. All line wavelengths are in vacuum and all magnitudes are expressed in the AB system \citep{Oke83}. For clarity, we refer to all reddening and SFR measurements made over the entire galaxy area as ``global'' measurements. 

\section{Data and Sample Selection}\label{sec:data}
%
%
\begin{figure*}
\includegraphics[width=\textwidth]{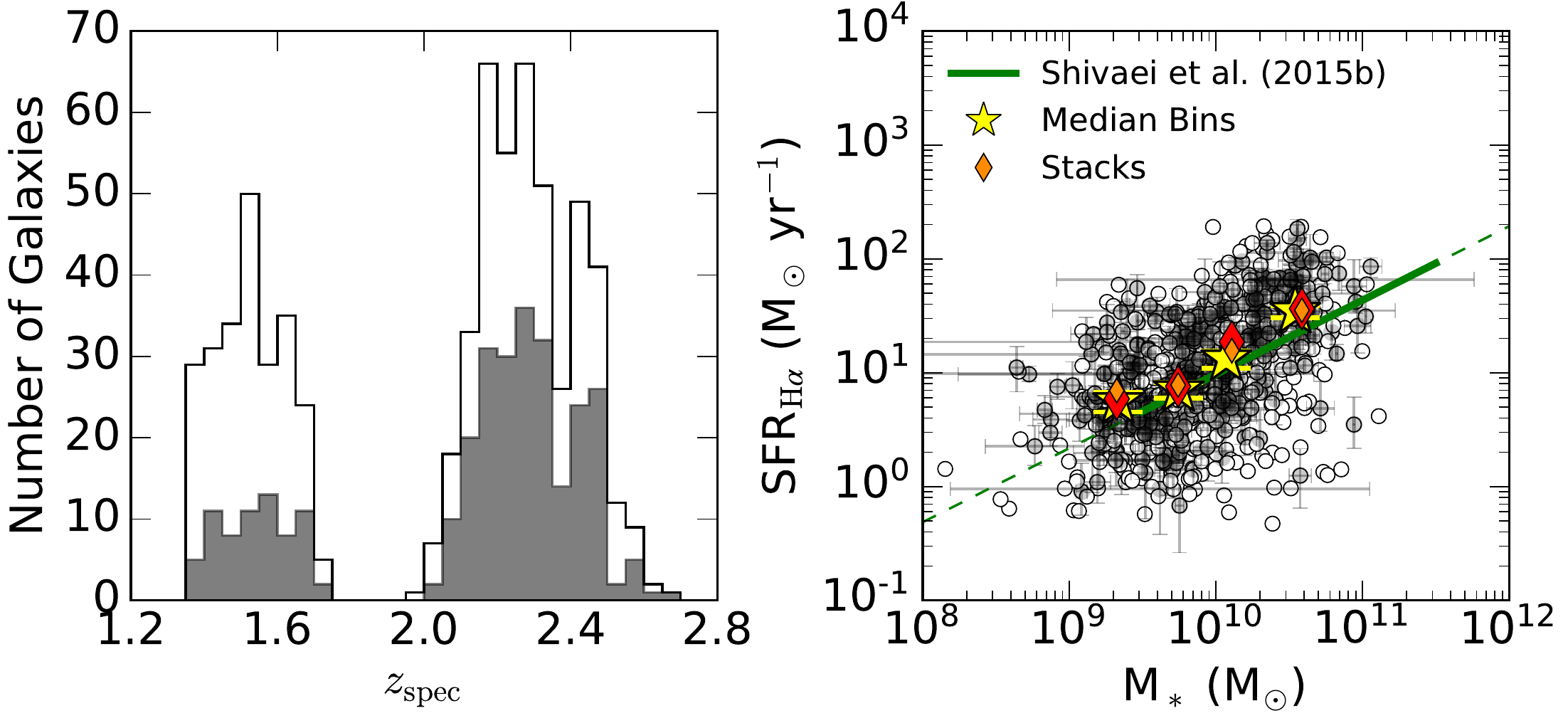}
\caption{\textit{Left:} The distribution of spectroscopic redshifts for the sample covering two redshift bins: $\izmin<z<1.74$ and $1.99<z<\izmax$. The spectroscopic redshifts of the initial sample (\insamp~galaxies) are shown by the white histogram. The final sample (\nsamp~galaxies), where all galaxies are detected with a $S/N\geq3$ in \HA\ and \HB\ emission, is denoted by the gray histogram. \textit{Right:} The sample relative to the star-forming \HA\ SFR--$M_*$ relation. \HA\ emission is corrected for dust reddening by using the Balmer decrement and assuming the \citet{Cardelli89} extinction curve. A \citet{Chabrier03} IMF and 20\% solar metallicity is also also assumed when calculating \HA\ SFRs (see \autoref{sec:ebvsfr}). The linear relation that is best-fit to the star-forming SFR--$M_*$ relation using the first two years of data from the MOSDEF survey \citep{Shivaei15} is shown by the solid green line and is extended with a dashed line for visual clarity. The MOSDEF spectra are stacked in four equal bins of stellar mass for both the initial (large red diamonds) and final (small orange diamonds) samples, respectively, showing that the sample is not biased against the dustiest galaxies where \HB\ is not detected. The galaxies in the initial (\insamp) and final (\nsamp) samples are shown by the empty and filled circles, respectively. The sample is also divided equally into four bins of stellar mass to show the medians of the individual measurements for the \nsamp\ galaxies in the sample (yellow stars).}
\label{fig:sample}
\end{figure*}

\subsection{CANDELS/3D-HST Photometry}\label{sec:3dhst}
CANDELS observed $\sim$900\,arcmin$^2$ of the sky and achieved 90\% completeness at $H_{160}\sim25$\,mag covering five well-studied extragalactic fields: AEGIS, COSMOS, GOODS-N, GOODS-S, and UDS \citep{Grogin11, Koekemoer11}. We use the multi-wavelength \HST/WFC3 (F125W, F140W, and F160W filters, hereafter $J_{125}$, $JH_{140}$, and $H_{160}$) and \HST/ACS (F435W, F606W, F775W, F814W, and F850LP filters, hereafter $B_{435}$, $V_{606}$, $i_{775}$, $I_{814}$, and $z_{850}$) resolved CANDELS imaging that has been processed by the 3D-HST grism survey team \citep{Brammer12, Skelton14, Momcheva16}. The publicly-available\footnote{\url{https://3dhst.research.yale.edu/}} processed CANDELS images have been drizzled to a 0\farcs06\,pix$^{-1}$ scale and PSF-convolved to the spatial resolution of the $H_{160}$ imaging (0\farcs18). We also use the broadband photometric catalogs provided by the 3D-HST team, which cover 0.3 to 8.0\,$\mu$m in the five extragalactic CANDELS fields. 

\subsection{MOSDEF Spectroscopy}\label{sec:mosdef}
Targets were selected for the MOSDEF survey in three redshift bins ($1.37<z<1.70$, $2.09<z<2.61$, and $2.95<z<3.80$) using the 3D-HST photometric and spectroscopic catalogs. In order to reach a stellar mass limit of $\sim$10$^9$\,\Msun~for all three redshift bins, an $H_{160}$-band limit of 24.0, 24.5, and 25.0\,mag was used for the $z\sim1.5$, $z\sim2.3$, and $z\sim3$ bins, respectively. These redshift bins were selected such that the following strong rest-frame optical emission lines fall in near-IR windows of atmospheric transmission: [OII]$\lambda3727,3730$, \HB, [OIII]$\lambda\lambda4960,5008$, \HA, [NII]$\lambda\lambda6550,6585$, and [SII]$\lambda6718,6733$.

Observations for the MOSDEF survey were taken over 48.5 nights in 2012--2016 on the 10\,m Keck I telescope using the MOSFIRE multi-object spectrograph \citep{McLean10, McLean12} in the $Y$, $J$, $H$, and $K$ bands ($R=3400$, 3000, 3650, and 3600 using 0\farcs7 slit widths). Rest-frame optical spectra were obtained for $\sim$1500 galaxies at $1.4\lesssim z\lesssim3.8$. Every slit mask included at least one slit star, which is used, in part, for the absolute flux calibration of the spectra and to correct for slit losses. The procedures for correcting for slit loss and measuring line fluxes are described in \autoref{sec:slitloss} and \autoref{sec:lineflux}, respectively. See \citet{Kriek15} for further details on the MOSDEF survey, including the observation strategy and data reduction. 

\subsubsection{Slit-loss Corrections}\label{sec:slitloss}
The slit loss estimated from the slit star is dependent on seeing conditions. Therefore, emission lines that are measured from different bandpasses (such as \HA\ and \HB) for a given galaxy will have different slit-loss corrections. Given that the typical seeing during the observations is comparable to the slit width of 0\farcs7, using a point source slit star is insufficient to correct the spatially-resolved MOSDEF galaxy targets for light lost outside of the spectroscopic slit. Therefore, to best account for additional slit loss, a secondary correction is applied to the spectrum of each galaxy in each near-IR filter. This correction is based on a 2D rotated elliptical Gaussian that is fit to the 3D-HST $H_{160}$ image of the galaxy that has been convolved to the typical seeing during the observation through the relevant filter \citep{Reddy15}. 

\subsubsection{Line Flux Measurements}\label{sec:lineflux}
Line fluxes are measured by fitting a linear function to the continuum and a Gaussian function to each emission line, with double and triple Gaussians fit to the [OII]$\lambda\lambda3727,3730$ doublet and \HA+[NII]$\lambda\lambda6550,6585$ lines, respectively. \HA\ and \HB\ line fluxes are corrected for underlying Balmer absorption using the stellar population model that best fits the observed 3D-HST broadband photometry (\autoref{sec:sedfit}). Flux errors are obtained by perturbing the 1D spectra by their error spectra 1000 times, remeasuring the line fluxes, and taking the 68th-percentile width of the distribution. Finally, the observed wavelength of the highest signal-to-noise ($S/N$) line, which is usually \HA\ or [OIII]$\lambda5008$, is used to measure the spectroscopic redshift. For more details on the emission line flux measurements, see \citet{Reddy15}.

\subsection{Sample Selection}\label{sec:sample}
The sample is derived from the MOSDEF parent sample in the $z\sim1.5$ and $z\sim2.3$ redshift bins, where both the \HA\ and \HB\ emission lines are covered by the ground-based spectroscopy. In order to optimally measure Balmer decrements and correct for slit loss, both \HA\ and \HB\ are required to be detected with a $S/N\geq3$. However, we also consider an ``initial'' sample where \HB\ is undetected ($S/N<3$), but still covered by the observations, to investigate a possible bias against the dustiest galaxies. AGNs are identified through their X-ray luminosities, optical emission lines ($\log{(\text{[NII]/H}\alpha)} > -0.3$), and/or mid-IR luminosities \citep{Coil15, Azadi17, Azadi18, Leung19}, then removed from the sample. 

With these requirements, the initial and primary samples include \insamp\ and \pnsamp~star-forming galaxies, respectively, at $\izmin<z<\izmax$. The primary sample includes 19 galaxies that fall outside of the targeted MOSDEF redshift bins, as these targets were serendipitously detected. When the CANDELS \HST\ images are smoothed to the seeing of the MOSFIRE observations (see \autoref{sec:slit_smooth}), 25 galaxies are undetected in the segmentation maps based on the smoothed images and are removed from the sample. After modeling the stellar populations globally and within the slit area (see \autoref{sec:sedfit}), we find that the SED-derived SFRs (i.e., UV SFRs) of 112 galaxies are measured to be higher inside the slit than across the entire galaxy---which is unphysical and primarily caused by key features of the SED being unconstrained (e.g., UV slope, Balmer/4000\,\AA\ breaks; see \autoref{sec:sedfit}). Therefore, these galaxies are removed from the sample along with an additional 9~galaxies that have poorly constrained photometry, resulting in a final sample size of \nsamp~star-forming galaxies at $\zmin<z<\zmax$.

\autoref{fig:sample} shows the spectroscopic redshift distribution and SFR--$M_*$ relation for the galaxies in the primary sample (\nsamp~galaxies; filled gray circles) compared to the initial sample (\insamp~galaxies; empty circles). As a visual aid, the sample is divided into four equally sized bins and their medians in SFR and stellar mass are shown (yellow stars). The values for the median bins are on average 0.14\,dex above the \citet{Shivaei15} SFR--$M_*$ relation, which was defined by the star-forming galaxies observed during the first two years of the MOSDEF survey. To ensure that we are not biasing the sample against the dustiest galaxies by removing all of those where \HB\ is undetected, stacks of the spectra are constructed using \texttt{specline}\footnote{\url{https://github.com/IreneShivaei/specline/}} \citep{Shivaei18} in four bins of stellar mass for the galaxies in the initial (\insamp\ galaxies; large red diamonds) and the final (\nsamp\ galaxies; small orange diamonds) samples. If our sample was biased against the dustiest galaxies, the stacks of the final sample (small orange diamonds) would be systematically higher than the stacks of the initial sample (large red diamonds) since \EBVg\ is used to correct \SFRha\ (see dust-corrected \SFRha\ in \autoref{sec:ebvsfr}). The right panel of \autoref{fig:sample} shows that the two sets of stacked bins are comparable within their uncertainties, are within the intrinsic scatter of the \citet{Shivaei15} SFR--$M_*$ relation, and the final sample (small orange diamonds) is not systematically biased towards higher \SFRha. Furthermore, stacking all of the \HB\ non-detections (226~galaxies) into a single bin reveals a Balmer Decrement measurement (\HA/\HB~$=4.13\pm0.74$) that is within $1\sigma$ of the average Balmer Decrement for the galaxies in our final sample (\HA/\HB~$=3.47\pm0.19$). Visual inspection of the spectra reveals that \HB\ non-detections for MOSDEF targets are often due to contamination from skylines rather than \HB\ falling below the detection threshold \citep{Reddy15, Shivaei15, Sanders18}. Thus, the sample is not significantly biased against the dustiest galaxies when \HB\ non-detections are excluded from the sample.

\section{Methodology and Measurements}\label{sec:methods}
\begin{figure}
\includegraphics[width=\columnwidth]{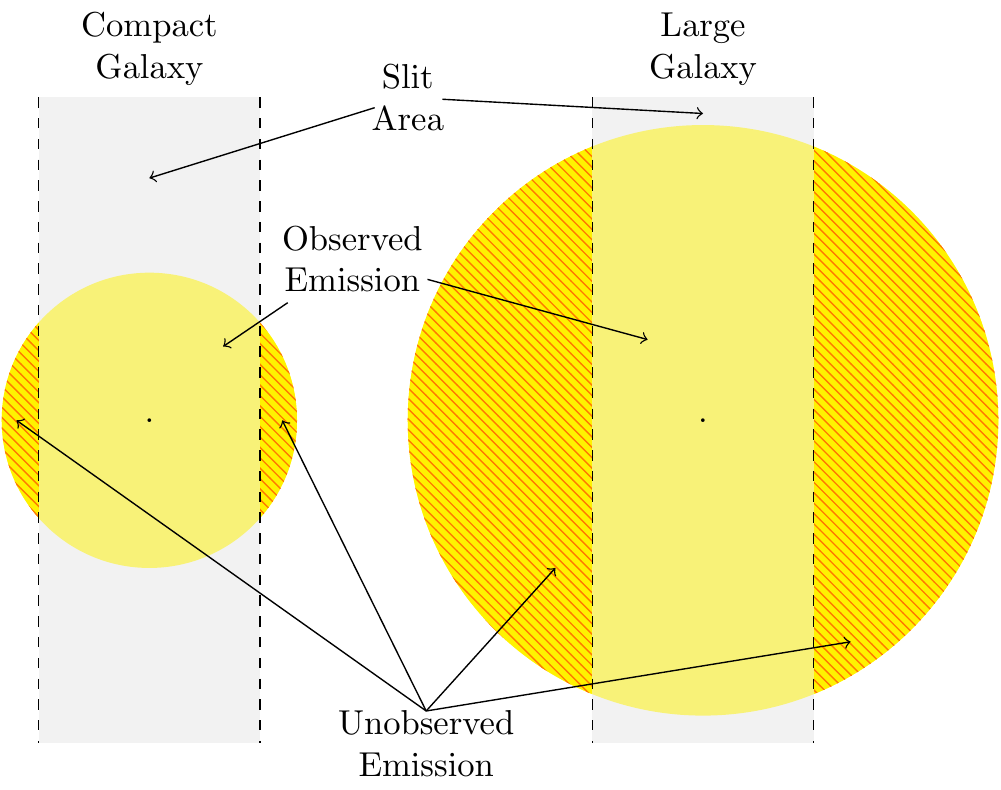}
\caption{Visualization of how the spectroscopic slit covers compact versus large galaxies. The gray region between the dashed vertical lines represents the spectroscopic slit area, which is centered (black dot) on the light distribution of each galaxy (yellow circles) measured from the photometry. Only the emission within the slit area is directly observed in the spectroscopy, whereas any emission that lies outside of the slit area (red hatched regions) is not directly observed.}
\label{fig:picture}
\end{figure}
In this section, we outline our procedure for calculating dust-corrected SFRs measured from \HA\ emission and the UV stellar continuum. \HA\ and \HB\ emission measured from slit spectroscopy (and other observed spectroscopic measurements) do not include information about the regions of the galaxy lying outside of the slit area, as is illustrated in \autoref{fig:picture}. The slit-loss corrections applied to the MOSDEF spectroscopic observations assume a certain light profile for the galaxy based on a single filter (see \autoref{sec:slitloss}). This simple light profile may not be sufficiently accurate for large galaxies where a significant fraction of their light falls outside the slit area or which may have light profiles that vary significantly from band-to-band. Furthermore, the position of the slit is randomly oriented relative to the physical structure of individual galaxies in order to optimize their placement on the MOSFIRE mask for multi-object spectroscopy. The UV light, however, is typically measured over an area larger than the spectroscopic slit, usually a circular or isophotal aperture in the imaging. Thus, to make a truly fair comparison between \HA\ and UV SFRs, we must measure them over the same spatial area, or aperture. 

First, we describe the image smoothing and flux measurements accounting for resolution and area in \autoref{sec:slit_smooth}. In \autoref{sec:sedfit}, we present the SED fitting procedure that is used to obtain both the global and slit-measured stellar population properties, from which \EBVs\ is directly inferred. Finally, in \autoref{sec:ebvsfr}, we detail our methods for obtaining \EBVg, \HA\ SFRs, and UV SFRs. 

\subsection{Image Smoothing}\label{sec:slit_smooth}
The seeing of the MOSFIRE observations was typically $\sim$0\farcs7. For comparison, the FWHM of the \HST\ $H_{160}$ PSF is $\sim$0\farcs18. Therefore, in order to directly compare measurements made from the resolved imaging with the spectroscopic measurements, the \HST\ images are first smoothed to match the typical seeing of the MOSFIRE observations. The \texttt{astropy.convolution} python package \citep{Astropy_collaboration13, Astropy_collaboration18} is used to convolve the \HST\ images with a 2D gaussian kernel with FWHM $\sigma_{\text{kernel}}=\sqrt{\sigma_{\text{MOSFIRE}}^2-\sigma_{\text{HST}}^2}$, where $\sigma_{\text{HST}}$ is the FWHM of the \HST\ PSF equal to 0\farcs18, and $\sigma_{\text{MOSFIRE}}$ is the FWHM of the seeing from the MOSFIRE observations. This smoothing procedure is repeated assuming $\sigma_{\text{MOSFIRE}}$ is 0\farcs5--0\farcs9 in 0\farcs1 increments, to account for variations in the seeing during the MOSFIRE observations. The smoothed images that are selected for an individual galaxy depends on the average seeing during the respective MOSFIRE observations ($\sigma_{\text{MOSFIRE}}$). RMS maps corresponding to the smoothed \HST\ images are constructed by scaling the original RMS maps such that the S/N is preserved in each pixel \citep[see][]{Fetherolf20}.

A segmentation map based on the smoothed images is generated by using a noise-equalized $J_{125}$+$JH_{140}$+$H_{160}$ smoothed image for source detection and following the same input parameters for Source Extractor \citep{Bertin96} as \citet{Skelton14}. There are 25\,galaxies from the sample that are undetected in the segmentation map of the smoothed images and, thus, are removed from the sample. The smoothed images that assume a $\sigma_{\text{MOSFIRE}}$ that best matches the MOSFIRE seeing from the observations of each galaxy are used to measure fluxes and perform SED fitting (see \autoref{sec:sedfit}).

Pixels that fall inside the slit area are identified using the source centroid, mask position angle, and slit width. The counts from the smoothed images for the pixels inside the slit are summed and converted to AB magnitude and the magnitude error is determined similarly (except summed in quadrature) from the noise maps. In order to avoid any single photometric point from skewing the SED fit, the minimum magnitude error is restricted to be no smaller than 0.05\,mag. 

To better constrain the shape of the SED at longer wavelengths, the resolved CANDELS imaging are supplemented with unresolved \textit{Spitzer}/IRAC photometry at 3.6\,$\mu$m, 4.5\,$\mu$m, 5.8\,$\mu$m, and 8.0\,$\mu$m available from the 3D-HST broadband photometry catalog \citep{Skelton14}. The \textit{Spitzer}/IRAC photometry has been corrected for contamination by neighboring sources using models of the \HST\ images that have been PSF-smoothed to the lower resolution IRAC photometry. Source pixels were identified from the segmentation maps. In order to include the IRAC photometry with the photometric slit measurements for SED fitting, the total IRAC fluxes from the 3D-HST broadband photometric catalog, $F_{\text{IRAC,tot}}$, are normalized using the $H_{160}$ flux measured inside the slit area from the smoothed imaging, $H_{160,\text{slit}}$, as follows:
\begin{equation}
\label{eq:irac}
F_{\text{IRAC,slit}}=F_{\text{IRAC,tot}}\frac{H_{160,\text{slit}}}{H_{160,\text{tot}}}
\text{,}
\end{equation}
where $H_{160,\text{tot}}$ is the total $H_{160}$ flux measured from the 3D-HST broadband photometry and $F_{\text{IRAC,slit}}$ is the resultant normalized IRAC flux. On scales smaller than the slit width \citet{Fetherolf20} showed that incorporating the IRAC photometry into the SED fitting using \autoref{eq:irac} does not significantly affect the derived stellar population properties.  

\subsection{SED Fitting}\label{sec:sedfit}
The total and slit-measured fluxes are modeled, as follows, to obtain SED-derived \EBVs, stellar population ages, SFRs, and stellar masses. First, the contribution of the strongest emission lines measured in the MOSFIRE spectra (i.e., [OII]$\lambda\lambda3727,3730$, \HB, [OIII]$\lambda\lambda4960,5008$, and \HA) is removed from the broadband photometry. We use $\chi^2$ minimization to select the \citet{Bruzual03} stellar population synthesis model that best fits the photometry assuming a \citet{Chabrier03} IMF, SMC extinction curve \citep{Fitzpatrick90, Gordon03}, and sub-solar metallicity (0.2\,$Z_{\odot}$).\footnote{An SMC-like or steeper extinction curve paired with sub-solar metallicities has been found to be more appropriate for young, high-redshift galaxies compared to the \citet{Calzetti00} attenuation curve \citep{Reddy18}. However, alternatively assuming the \citet{Calzetti00} attenuation curve and solar metallicities primarily affects the absolute mass measurements and not their relative order \citep{Reddy18}. Therefore, the results presented here do not significantly change if we were to alter the assumed attenuation curve.} Reddening is allowed to vary in the range $0.0 \le \EBVs\ \le 0.4$. Only constant SFHs and stellar ages between 50\,Myr and the age of the Universe at the redshift of each galaxy are considered.\footnote{The SFRs of $z\sim2$ galaxies are best reproduced using either exponentially rising or constant SFHs when the stellar population ages are restricted to being older than the typical dynamical timescale \citep[50\,Myr;][]{Reddy12-1}. Assuming constant SFHs typically produces stellar population ages that are $\sim$30\%\ younger than those derived from exponentially rising SFHs \citep{Reddy12-1}. The UV SFRs (see \autoref{sec:ebvsfr}) for galaxies in our sample are on average 0.02\,dex higher when alternatively assuming exponentially rising SFHs, which is comparable with the typical uncertainty in \SFRuv. Therefore, alternatively assuming exponentially rising SFHs, or allowing SFH to be a free parameter, does not alter the conclusions presented in this work.} Finally, the measured photometric fluxes are perturbed 100 times by the flux errors and the $1\sigma$ SED parameter uncertainties are determined by the range of parameters inferred for the 68 models with the lowest $\chi^2$. 

There are 112~galaxies that exhibit SED-derived SFRs that are unphysically larger when measured inside the slit area than across the entire galaxy. We consider the SFRs measured inside the slit to be inaccurate in these cases considering that they are based on significantly fewer and less precise photometric measurements. \citet{Fetherolf20} showed that if the photometry does not constrain key features of the SED (e.g., UV slope, Balmer/4000\,\AA\ break) either by wavelength coverage or sufficient S/N, then the SED-derived parameters may be biased towards redder \EBVs\ and younger stellar population ages while stellar masses remain robust. Overestimated reddening and underestimated stellar ages (at constant stellar mass and SFH) can, in turn, lead to underestimated SFRs. All 112~galaxies that exhibit overestimated SED-derived SFRs inside the slit exhibit \EBVs\ and stellar ages that are both redder and younger than the globally averaged \EBVs\ and stellar ages. On the other hand, the average differences between global and slit-measured \EBVs\ and $\log(\text{Age/yr})$ for the other 312~galaxies are only 0.03\,mag and 0.09\,dex, respectively. Furthermore, 98 of the 112~galaxies with overestimated slit-measured SED SFRs have less than two photometric points with a $S/N\geq1$ covering the UV slope (1250--2500\,\AA),\footnote{There is always at least one filter covering the UV slope, but that filter does not necessarily have a $S/N\geq1$.} and thus this region of the SED is not well-constrained for these galaxies. There are also an additional 9~galaxies that do not have well-constrained photometric measurements from the resolved imaging. For these reasons, we remove from the sample a total of 121~galaxies that do not have robust constraints on the UV slope.

\subsection{\EBVg\ and SFR Calculations}\label{sec:ebvsfr}
\begin{figure*}
\begin{adjustbox}{width=\linewidth, center}
\begin{minipage}{\linewidth}
\includegraphics[width=.9\linewidth]{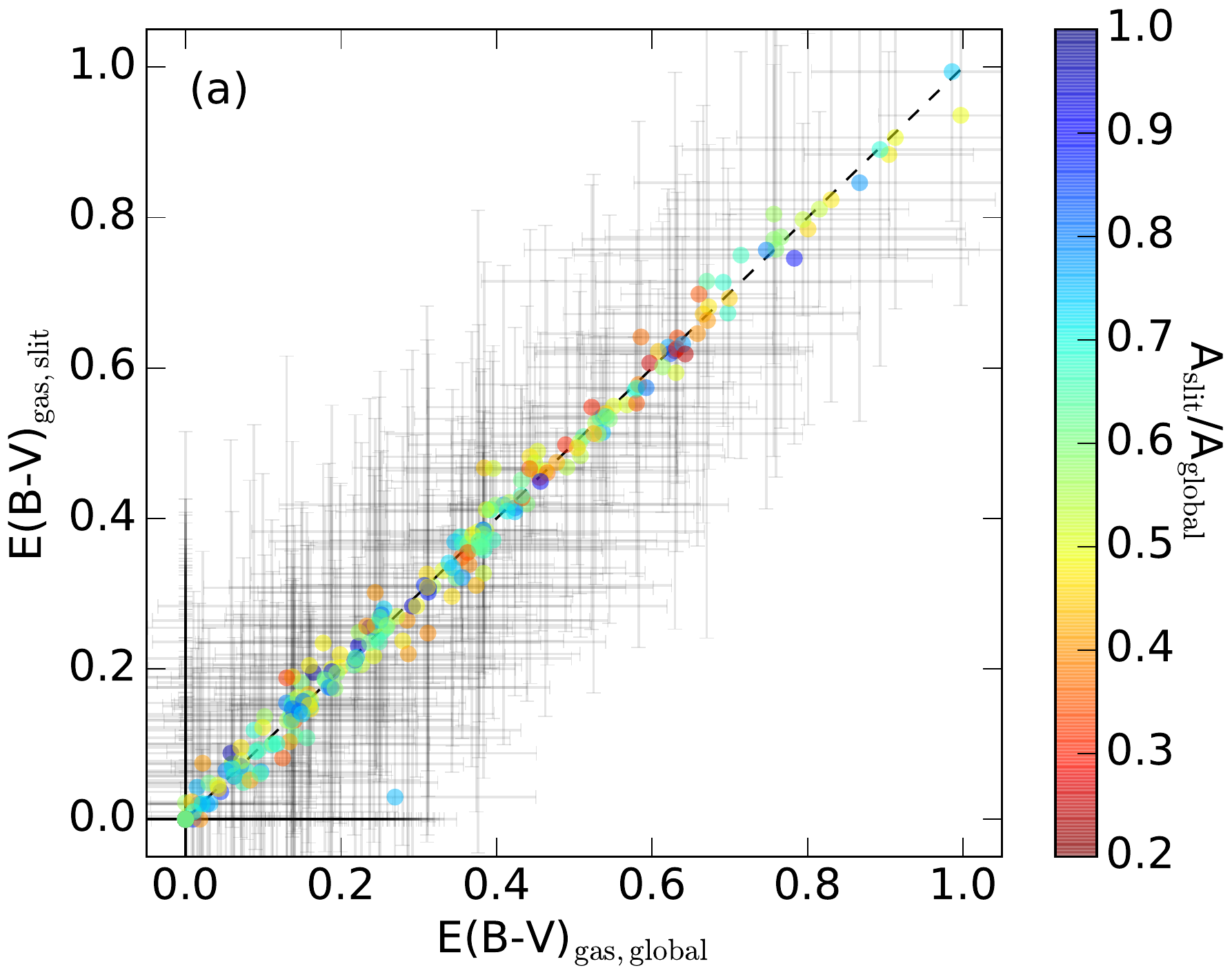}
\end{minipage}
\quad
\begin{minipage}{\linewidth}
\includegraphics[width=.9\linewidth]{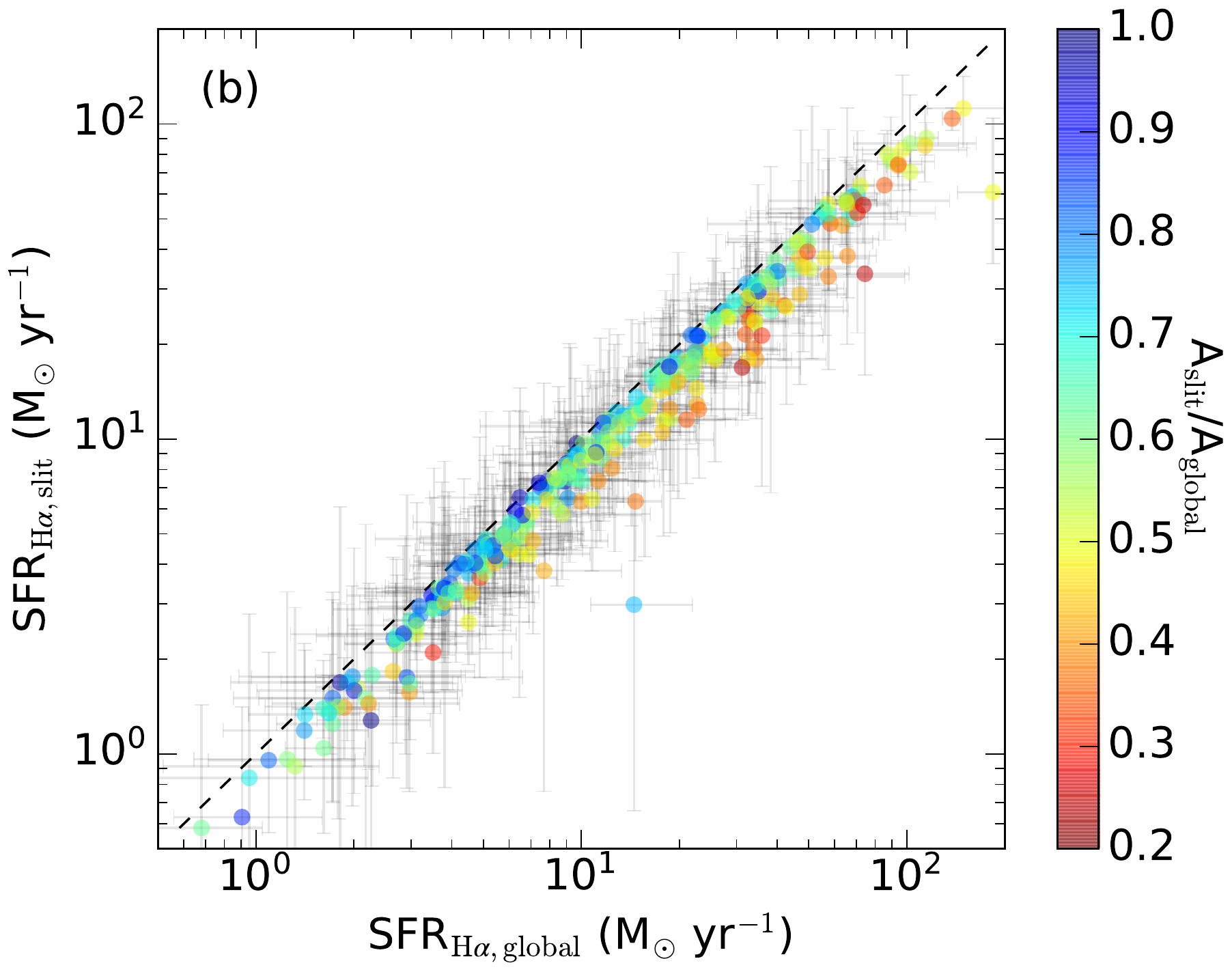}
\end{minipage}
\end{adjustbox}
\begin{adjustbox}{width=\linewidth, center}
\begin{minipage}{\linewidth}
\includegraphics[width=.9\linewidth]{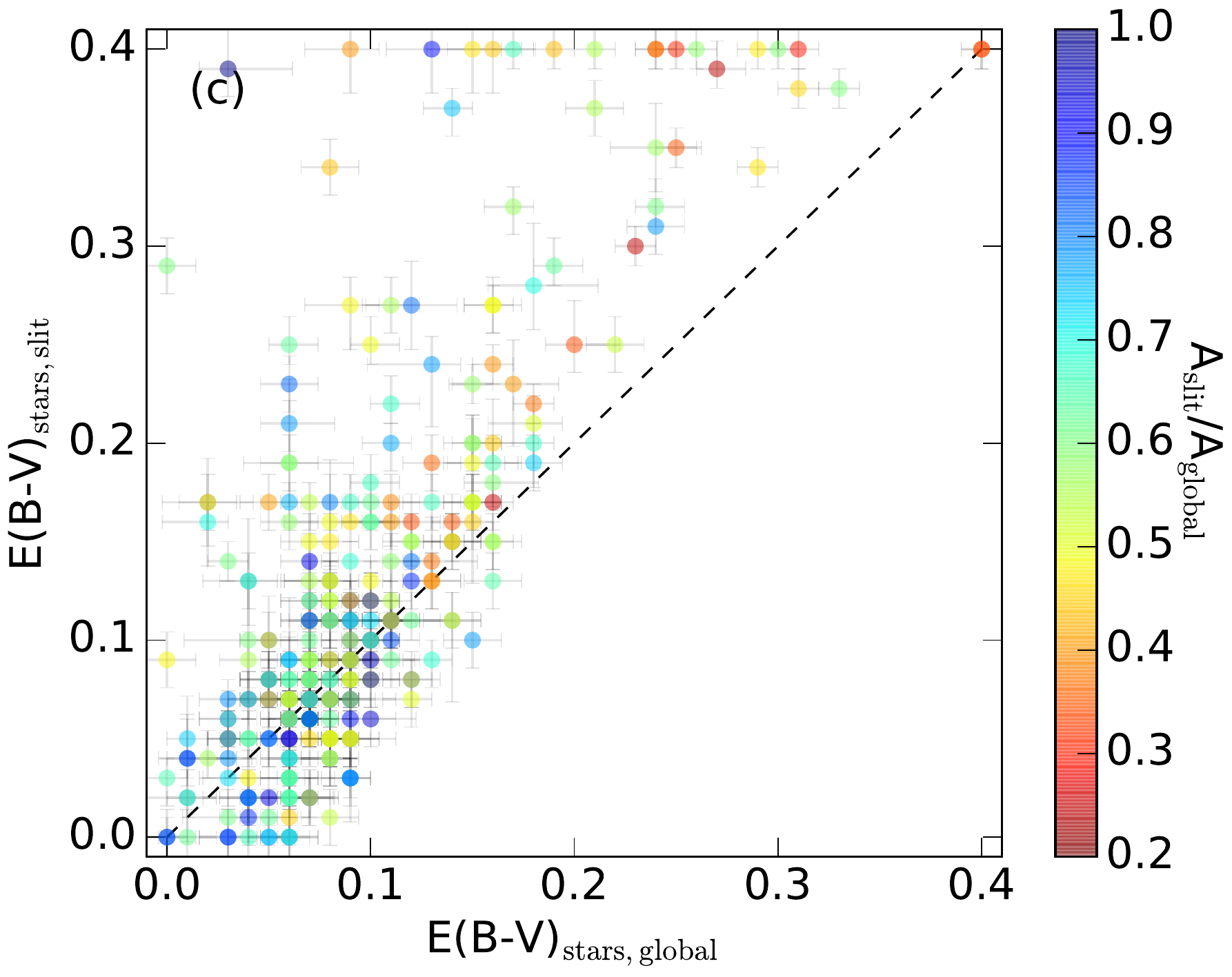}
\end{minipage}
\quad
\begin{minipage}{\linewidth}
\includegraphics[width=.9\linewidth]{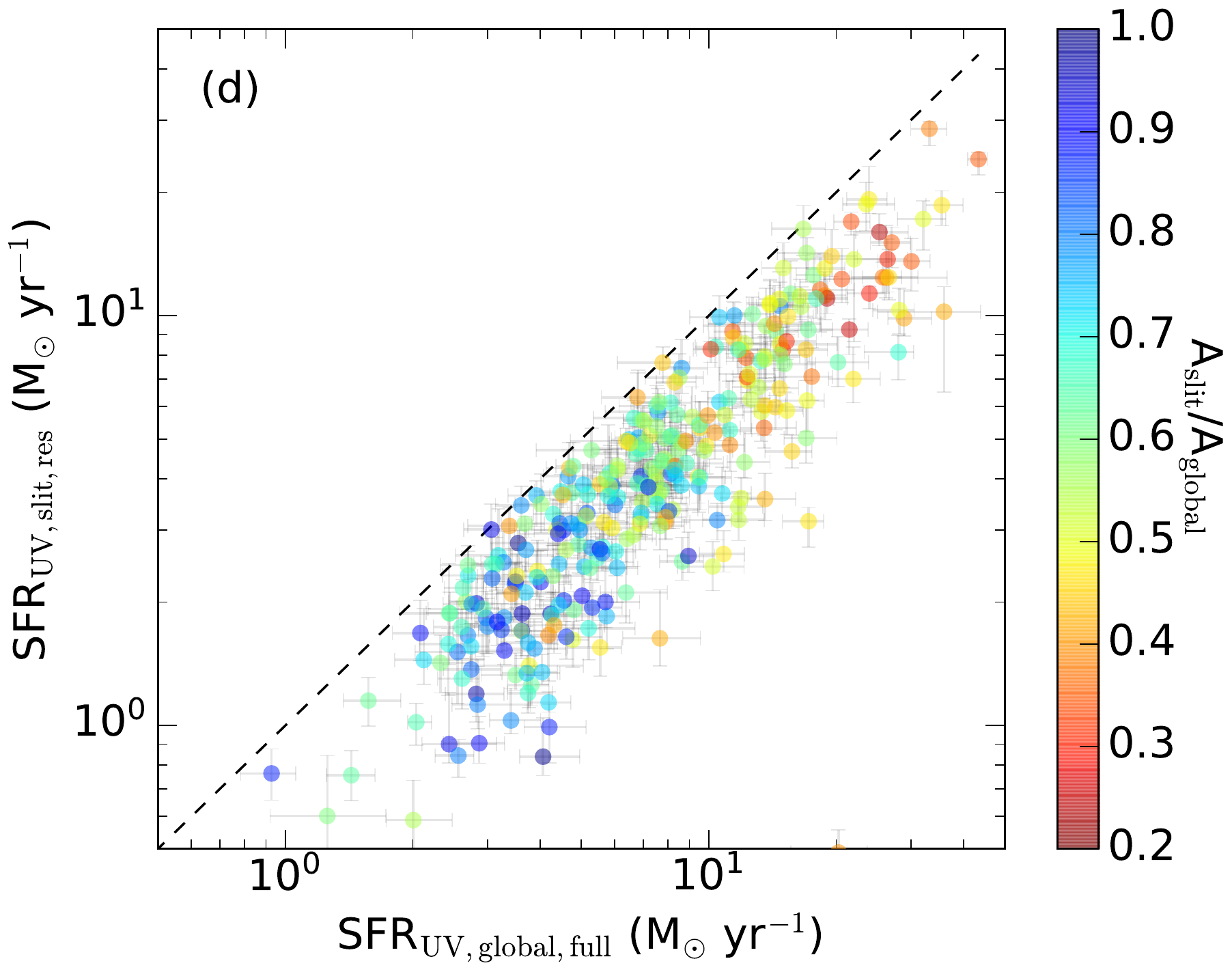}
\end{minipage}
\end{adjustbox}
\caption{\textit{Left:} Nebular (panel a) and stellar continuum (panel c) reddening obtained from the spectroscopic slit area compared to the inferred global reddening. \textit{Right:} \SFRha\ (panel b) and \SFRuv\ (panel d) derived from the slit area versus the global SFRs. The points are colored by the fraction of the galaxy that falls within the slit area (see \autoref{fig:picture}). The black dashed lines show where the global and slit-derived reddening and SFRs are equal.}
\label{fig:totslit_compare}
\end{figure*}
\begin{figure*}
\begin{adjustbox}{width=\linewidth, center}
\begin{minipage}{\linewidth}
\includegraphics[width=.9\linewidth]{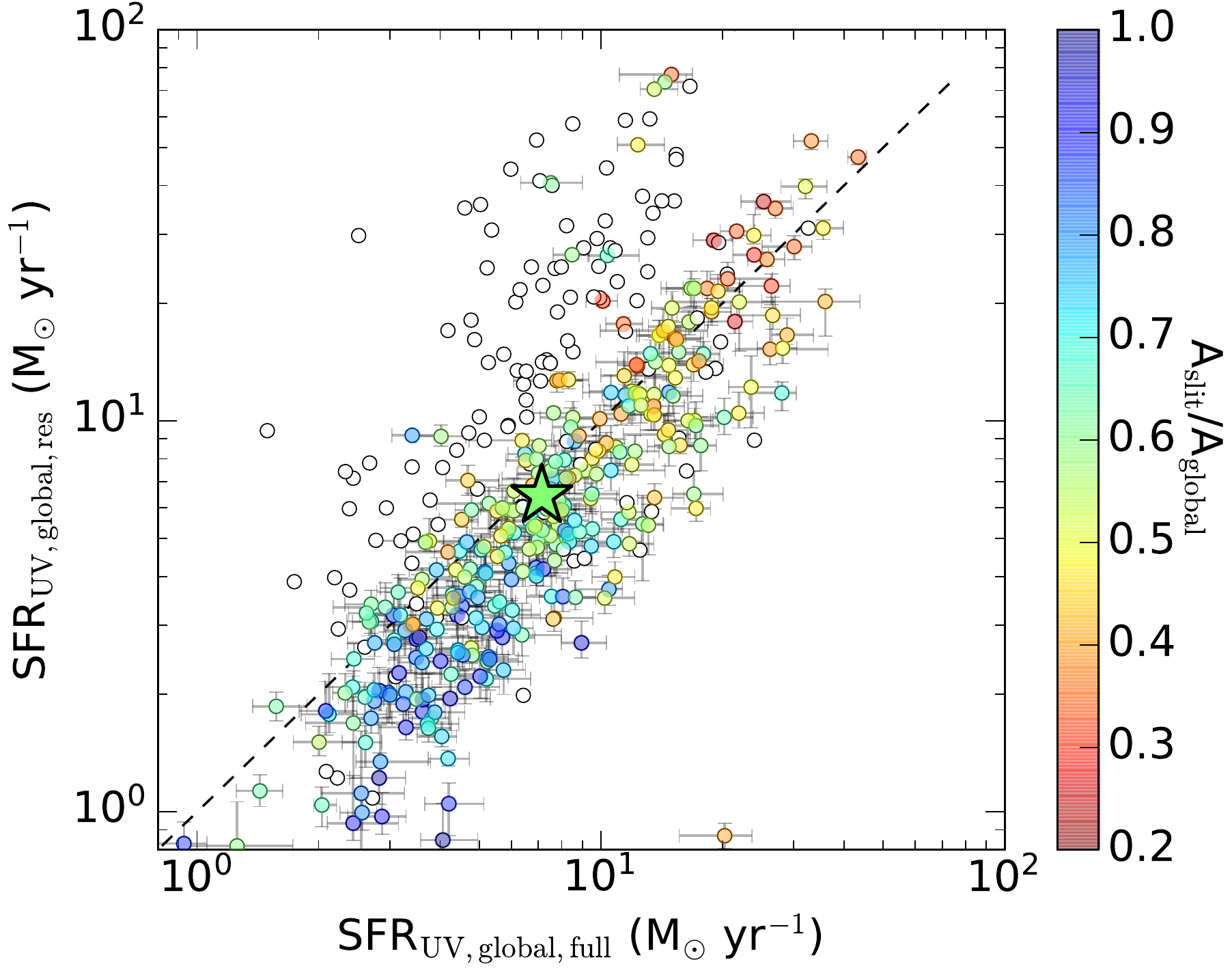}
\end{minipage}
\quad
\begin{minipage}{\linewidth}
\includegraphics[width=.9\linewidth]{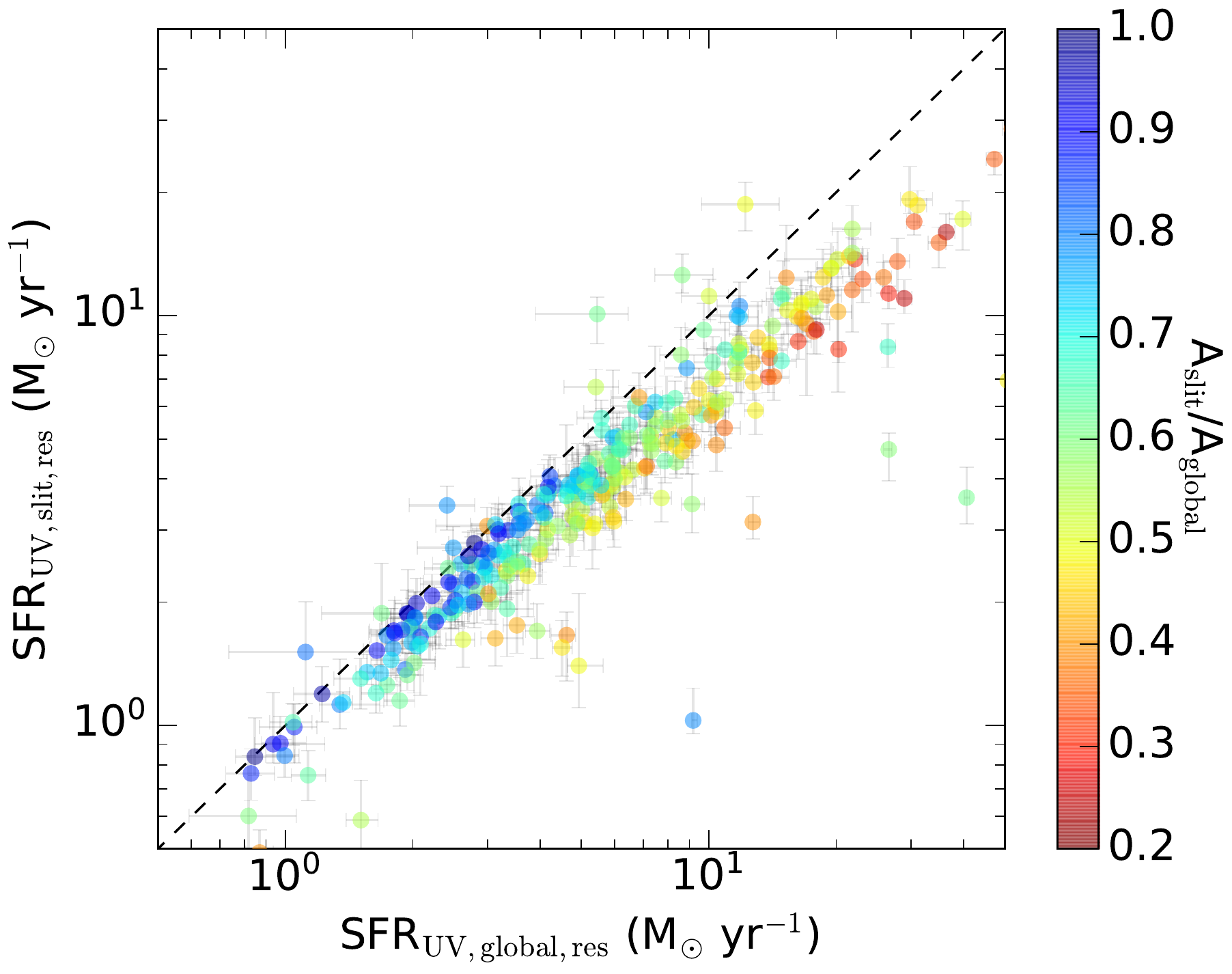}
\end{minipage}
\end{adjustbox}
\caption{\textit{Left:} \SFRuvtot\ derived from an SED fit to 3D-HST photometric catalog, which has extended wavelength coverage, compared to those derived from the summed flux of only the resolved CANDELS/3D-HST imaging. The dashed line shows where the two \SFRuv\ measurements are equal. The points are colored by the fraction of the galaxy that falls within the slit area (see \autoref{fig:picture}). The white empty points show the 121~galaxies that were removed from the sample due to the \SFRuvslit\ being larger than the \SFRuvtot\ or from having poorly constrained resolved photometry (see \autoref{sec:sedfit}). \textit{Right:} Same as \autoref{fig:totslit_compare}d, except that the \SFRuvtot\ is derived from an SED that is fit only to the resolved CANDELS/3D-HST imaging and normalized IRAC photometry, which is comparable to how the \SFRuvslit\ is derived exclusively within the slit area (see \autoref{sec:slit_smooth}).}
\label{fig:bbintSFR}
\end{figure*}
The globally-measured nebular reddening, \EBVg, is calculated by using the measured Balmer decrement (\HA/\HB) of each galaxy and assuming the \citet{Cardelli89} Galactic extinction curve:\footnote{Recently, \citet{Reddy20} directly measured the high-redshift nebular dust attenuation curve using the full MOSDEF sample and found that its shape is similar to the \citet{Cardelli89} Galactic extinction curve at rest-frame optical wavelengths.}
\begin{equation}
\label{eq:ebvgas}
\EBV_{\text{gas}} = \frac{2.5}{k(\HB)-k(\HA)}\log_{10}\left(\frac{\HA/\HB}{2.86}\right)\text{,}
\end{equation}
where $k(\HA)$ and $k(\HB)$ are extinction coefficients at \HA\ and \HB\ (2.53 and 3.61, respectively, assuming \citealt{Cardelli89} extinction).
\autoref{eq:ebvgas} assumes typical ISM conditions, namely Case B recombination, $T=10000$\,K, and $n_e=10^2$\,cm$^{-3}$, where the intrinsic ratio $\HA/\HB=2.86$ \citep{Osterbrock89}. The globally-measured dust-corrected \HA\ luminosity is then obtained using $k(\HA)$ from the \citet{Cardelli89} Galactic extinction curve and the \EBVg\ derived in \autoref{eq:ebvgas}.

In this study, we are interested in directly comparing UV (i.e., SED-derived) SFRs with spectroscopic dust-corrected \HA\ SFRs. Typically, \HA\ luminosities are converted to SFRs using the \citet{Kennicutt98-1} relation. However, the \citet{Kennicutt98-1} relation assumes solar metallicities, whereas we assume 20\% solar metallicities in the SED fitting (see \autoref{sec:sedfit}). To ensure that the \HA\ SFRs are consistent with the assumptions used to infer the SED-derived SFRs, we obtain an appropriate \HA\ luminosity-to-SFR conversion for each galaxy directly from the \citet{Bruzual03} SED model templates. For each individual galaxy, the best-fit stellar population age derived from the broadband photometry is used to select the appropriate SED model template. From the template, the rate of ionizing photons is measured by integrating the spectrum in the range $90\le\lambda\le912$\,\AA. The conversion from luminosity to SFR is derived using the effective recombination rate of Hydrogen atoms and the probability of \HA\ photons being emitted during recombination. Typical ISM conditions are again assumed for Case B recombination, $T=10000$\,K, and $n_e=10^2$\,cm$^{-3}$ \citep{Osterbrock89}. Deriving \HA\ SFRs using the SED model templates results in SFRs that are $\sim$24\% ($\sim$0.11\,dex) lower than those derived using the typically-assumed \citet{Kennicutt98-1} relation. All \HA\ SFRs throughout this paper, including those used to establish the \citet{Shivaei15} SFR--$M_*$ relation in \autoref{fig:sample}, have been corrected to assume the appropriate luminosity-to-SFR conversion derived here. Finally, the \EBVg\ and \HA\ SFRs corresponding to the slit aperture are computed by reversing the slit-loss corrections applied to \HA\ and \HB\ (see \autoref{sec:slitloss}) and repeating the above procedure.

For consistency with the way in which \HA\ SFRs are calculated, UV SFRs are also calculated directly from the best-fit \citet{Bruzual03} SED model template by taking the 1600\,\AA\ luminosity measured from the template and correcting it for dust assuming the SMC extinction curve \citep{Fitzpatrick90, Gordon03} and the SED-derived \EBVs. UV SFRs that are derived in this way are on average $\sim$3\% lower ($\sim$0.012\,dex) than the SED-derived SFRs. 

\section{Photometric vs. Spectroscopic Properties}\label{sec:slit}
The SFR is fundamental to understanding how galaxies assemble their stellar mass.  An important question is whether SFRs measured at different wavelengths agree as discrepancies may give clues about their sensitivity to star formation on different timescales \citep{Reddy12-1, Price14}, provide useful information about the IMF and SFH of galaxies \citep{Madau14}, or reveal how dust is distributed amongst various stellar populations \citep{Boquien09, Boquien15, Hao11, Reddy15, Katsianis17}. Alternatively, the assumptions made when correcting spectroscopic observations for light lost outside of the slit area may cause spectroscopic SFR measurements to disagree with photometric ones \citep{Brinchmann04, Kewley05, Salim07, Richards16, Green17}. In this section, we investigate whether discrepancies between \HA\ and UV SFRs persist when accounting for the different apertures over which the two are measured. 

First, in \autoref{sec:slittot_compare} we directly compare measurements made within the slit area with those inferred globally for \EBVg, \EBVs, \SFRha, and \SFRuv. We investigate how the difference between nebular and stellar continuum reddening varies as a function of SFR when all measurements are made directly within the slit in \autoref{sec:EBVdiff}. Similarly, in \autoref{sec:slitfrac}, we examine how the slit-measured \HA-to-UV SFR ratio varies as a function of slit-measured SFR. Finally, we consider the role of galaxy size in shaping the aforementioned relationships in \autoref{sec:areatrends}.

\subsection{Global vs. Slit Properties}\label{sec:slittot_compare}
\begin{figure*}
\begin{adjustbox}{width=\linewidth, center}
\begin{minipage}{\linewidth}
\includegraphics[width=\linewidth]{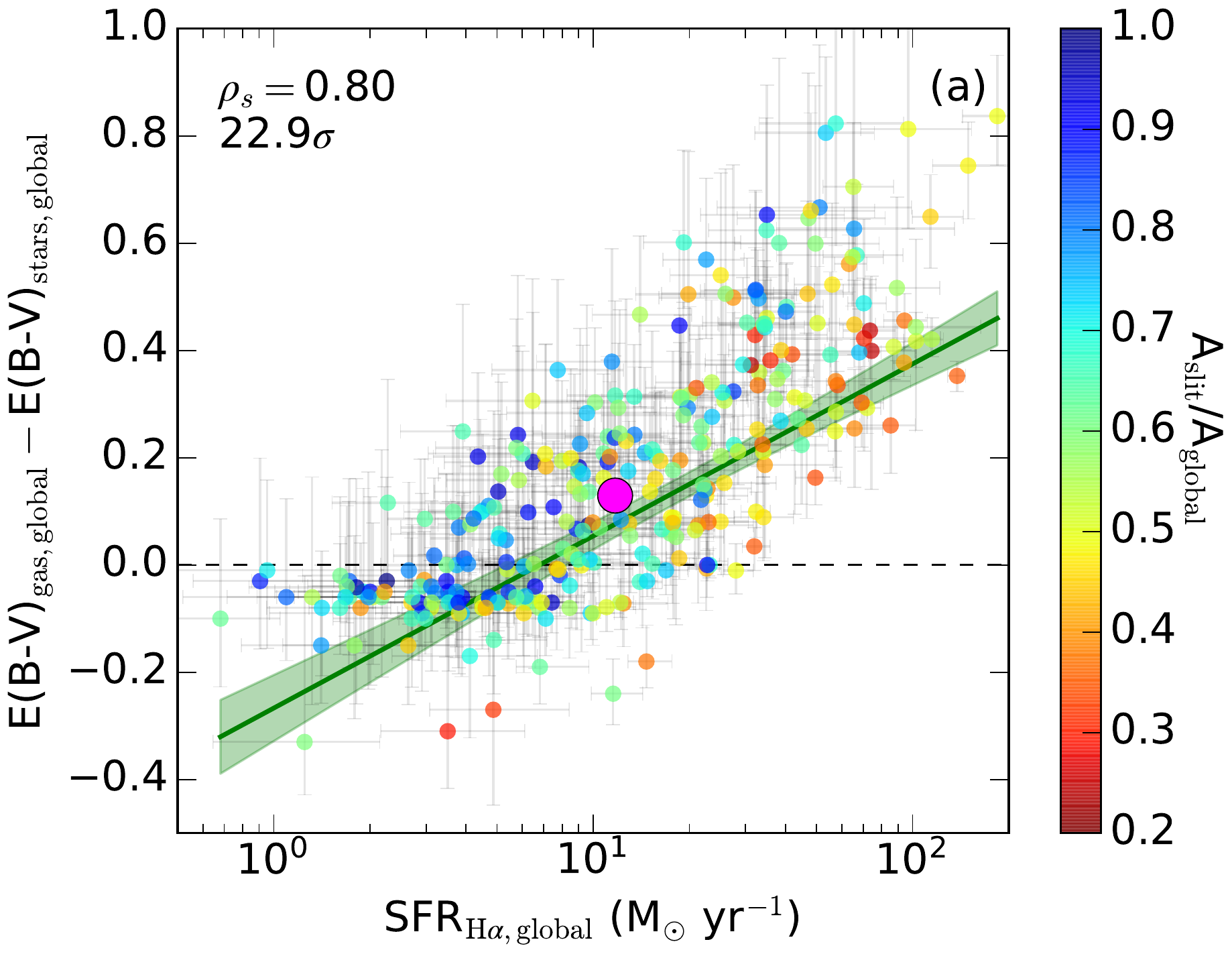}
\end{minipage}
\quad
\begin{minipage}{\linewidth}
\includegraphics[width=\linewidth]{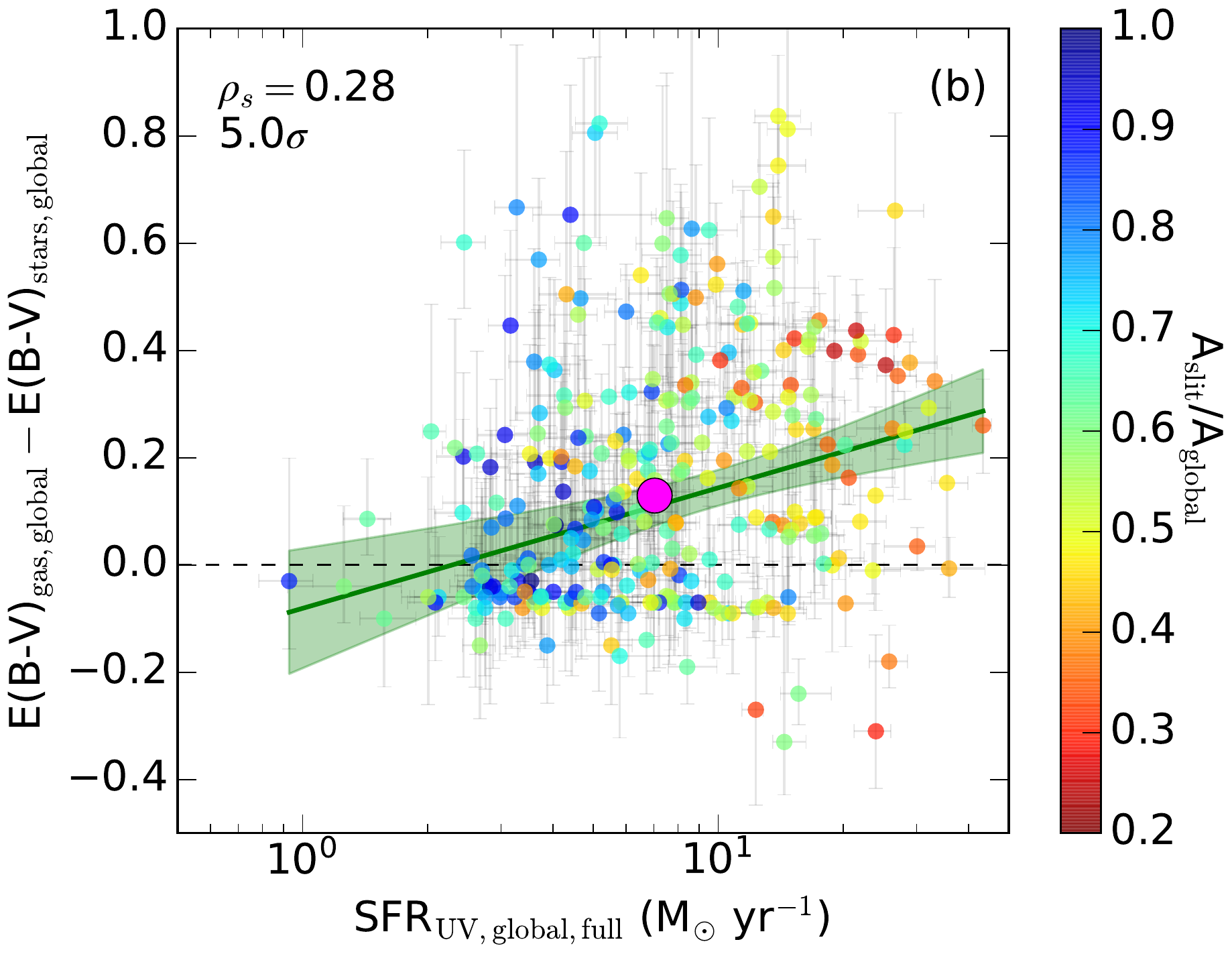}
\end{minipage}
\end{adjustbox}
\begin{adjustbox}{width=\linewidth, center}
\begin{minipage}{\linewidth}
\includegraphics[width=\linewidth]{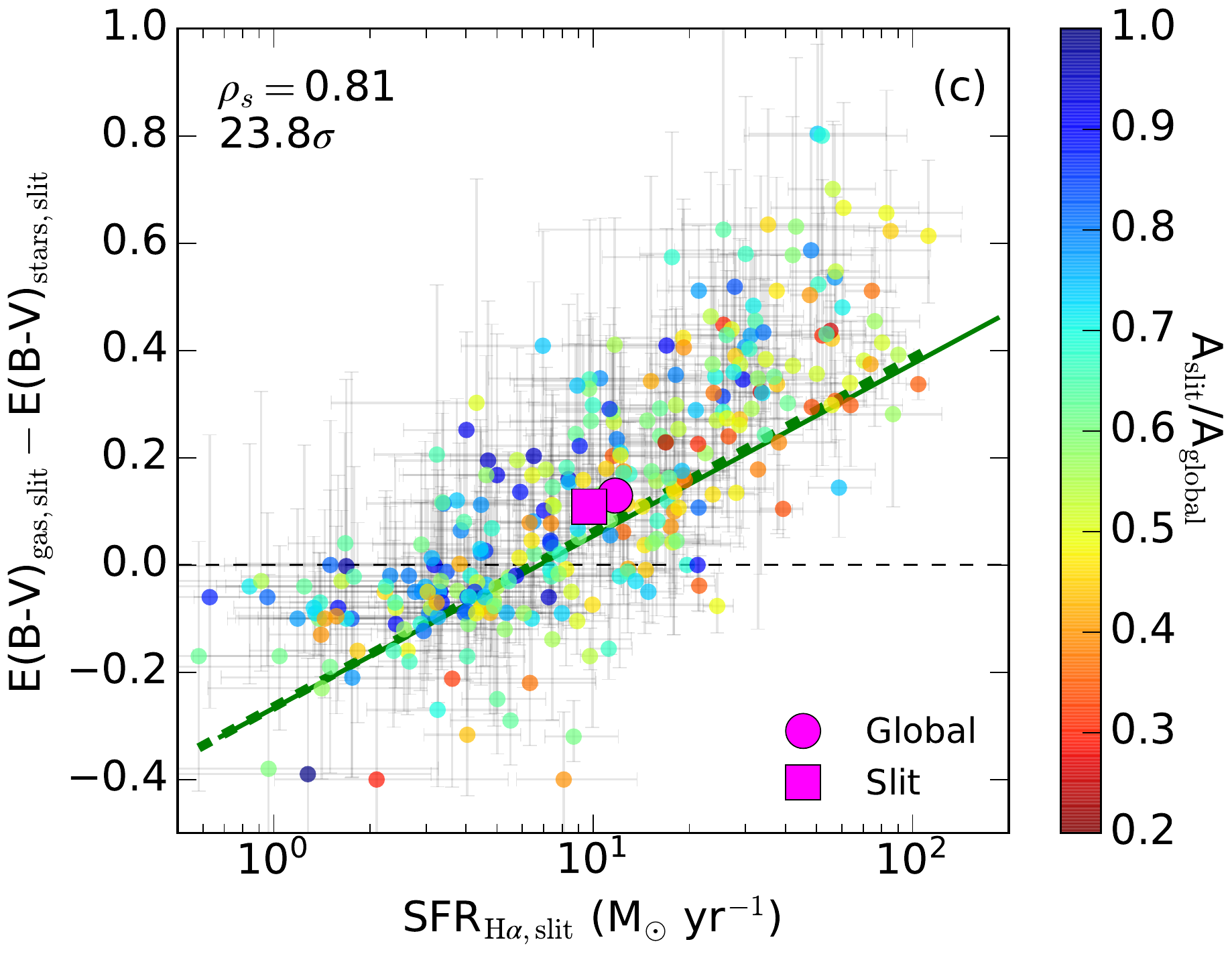}
\end{minipage}
\quad
\begin{minipage}{\linewidth}
\includegraphics[width=\linewidth]{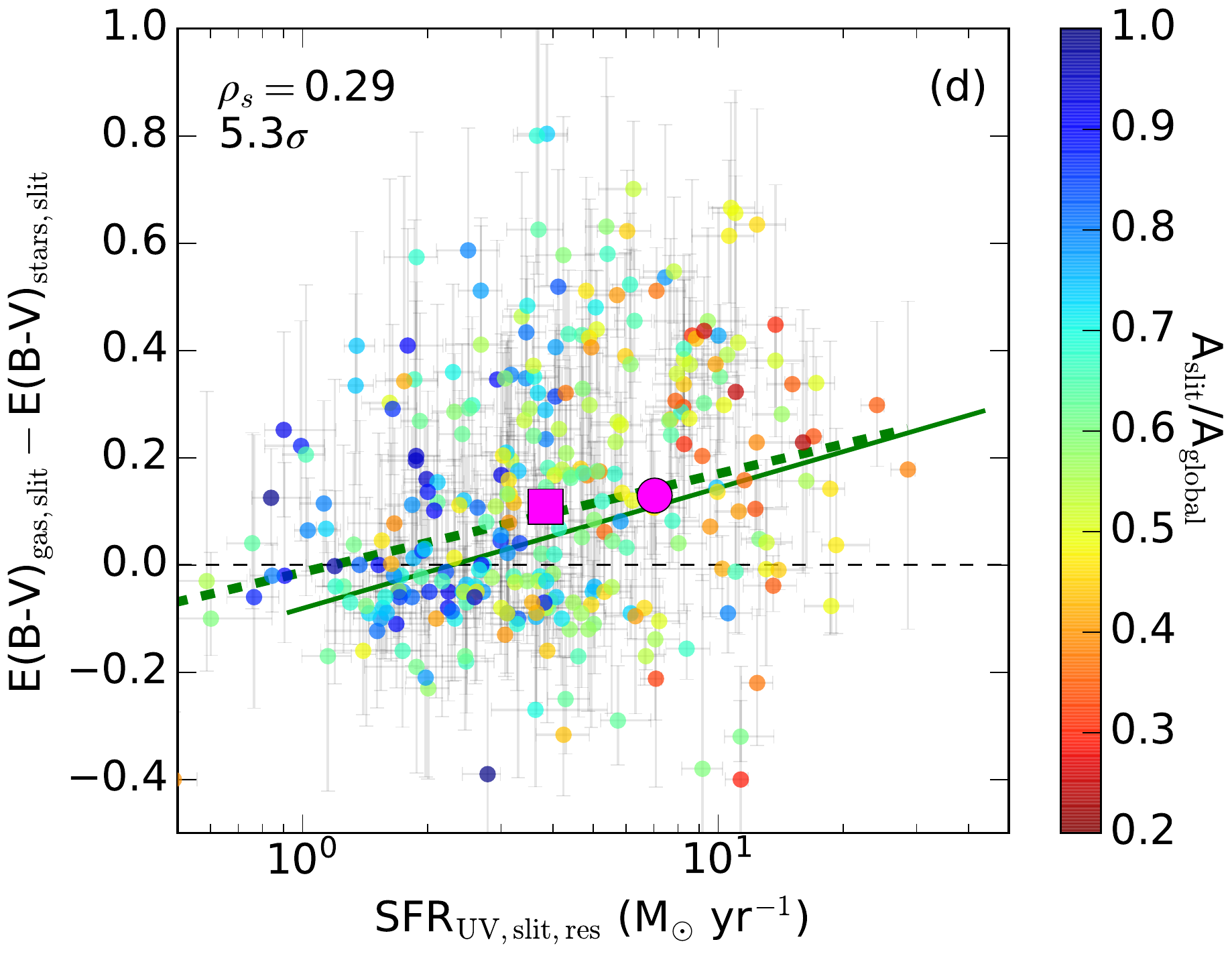}
\end{minipage}
\end{adjustbox}
\caption{Differences between nebular and stellar continuum \EBV\ versus \SFRha\ (\textit{left}) and \SFRuv\ (\textit{right}) for when all measurements are made globally (\textit{top}) or exclusively within the spectroscopic slit area (\textit{bottom}) for the \nsamp\ galaxies in the sample. The points are colored by the fraction of the galaxy that falls within the slit area (see \autoref{fig:picture}). The purple symbols show the median difference between \EBVg\ and \EBVs\ and the median \SFRha\ or \SFRuv\ for when measurements are made globally (circles) or directly within the slit area (squares). The horizontal black dashed lines shows where \EBVg\ and \EBVs\ are equal. The solid green lines show the best-fit linear relationship when all measurements are made globally and the dashed green lines show the best-fit linear relationship when all measurements are made directly within the slit region. The green shaded regions show the 3$\sigma$ confidence intervals of the best-fit linear relationships for the global measurements. The Spearman rank correlation coefficient and its significance are listed in the top left corner of each panel.}
\label{fig:EBVdiff_compare}
\end{figure*}
In \autoref{fig:totslit_compare}, the globally inferred nebular and stellar continuum reddening (left panels) and \HA\ and UV SFRs (right panels) are compared to those derived exclusively from the flux in the spectroscopic slit area. The points are color-coded by the fraction of the galaxy falling in the slit, \slitfrac, where A$_{\text{global}}$ is the are subsumed by the pixels that cover the surface area of the galaxy as identified by the segmentation map provided with the 3D-HST imaging, and A$_{\text{slit}}$ is the area of the subset of pixels that cover the same area as the MOSFIRE spectroscopic slit. Therefore, small values of \slitfrac\ represent physically large galaxies (red in color) and larger values of \slitfrac\ approaching unity represent compact galaxies (blue in color). Panels (a) and (b) of \autoref{fig:totslit_compare} compare the spectroscopic slit-based measurements with the global measurements, where the latter is inferred based on applying the slit-loss corrections discussed in \autoref{sec:slitloss}. The \EBVg\ determined from \HA\ and \HB\ emission is assumed to apply over the entire galaxy, and the slit-loss corrections assume that the \HA\ emission profile has the same shape as the optical light profile (i.e., $H_{160}$ image; see \autoref{sec:slitloss}). As a result, the slit-measured nebular reddening, \EBVgslit, is essentially identical to the global value, \EBVgtot\ (panel a), with only small deviations being attributed to minor differences between the slit-loss corrections for \HA\ and \HB\ caused by differences in seeing conditions of the observations of the relevant bands (see \autoref{sec:slitloss}). Naturally, \SFRha\ measurements within the slit, \SFRhaslit, are lower those measured globally, \SFRhatot\ (0.10\,dex lower on average), where the difference depends on the fraction of the galaxy falling in the slit (\slitfrac; panel b). Panel (c) shows that the reddening of the stellar continuum measured from the 3D-HST photometry, \EBVstot, is similar to that measured when only considering flux that falls in the slit, \EBVsslit, for most galaxies. There are 29~galaxies with \EBVsslit\ that are $>$2$\sigma$ ($\sim$0.12\,mag) redder (higher) than \EBVstot, which can be attributed to the UV slope being poorly constrained by the resolved photometry \citep[see][]{Fetherolf20}. As discussed in \autoref{sec:sample}, we required \SFRuv\ measured within the slit, \SFRuvslit, to be lower than \SFRuv\ measured globally, \SFRuvtot, in order to be included in the sample. Panel (d) shows that \SFRuvslit\ is on average 0.27\,dex lower than \SFRuvtot, and the size of the galaxy (i.e., \slitfrac) correlates more significantly with \SFRuvtot\ than the difference between slit and global \SFRuv\ measurements (see \autoref{sec:areatrends} for further discussion). 

The \SFRuvtot\ shown in \autoref{fig:totslit_compare}d is derived using the extended wavelength coverage available through the photometry provided by the 3D-HST photometric catalog, and includes unresolved photometry from other ground- and space-based facilities. The \SFRuvslit, on the other hand, only utilizes the photometry with resolved CANDELS/3D-HST imaging and normalized IRAC photometry (see \autoref{sec:slit_smooth}). In the left panel of \autoref{fig:bbintSFR}, these two methods for measuring \SFRuvtot\ are compared. The 121~galaxies with poorly constrained resolved photometry or with \SFRuvslit\ measurements that are higher than \SFRuvtot\ (white empty points; see \autoref{sec:sedfit}) are included to demonstrate that these ``unphysical'' measurements lie on the higher SFR side of the resultant scatter when \SFRuv\ are measured from fewer photometric points. It can be seen that \SFRuv\ measured from the 3D-HST photometric catalog agrees on average with those derived from only the resolved photometry (colored star symbol). 

The right panel of \autoref{fig:bbintSFR} is similar to \autoref{fig:totslit_compare}d, except that both \SFRuvtot\ and \SFRuvslit\ are derived only using the resolved CANDELS/3D-HST and normalized IRAC photometry. The right panel of \autoref{fig:bbintSFR} more clearly shows how the difference between \SFRuvtot\ and \SFRuvslit\ changes with galaxy size (i.e., \slitfrac) and that \SFRuvslit\ is 0.10\,dex lower on average than \SFRuvtot. As previously stated in \autoref{sec:sedfit}, \citet{Fetherolf20} showed that limited photometric coverage of the UV slope and other critical SED features (e.g., Balmer/4000\,\AA\ break) leads to redder \EBVs\ and younger ages. While measurements for \SFRuvslit\ can only utilize the wavebands with resolved imaging available (and normalized IRAC photometry), \SFRuvtot\ can be measured more robustly by using the data from several unresolved filters that is available in the 3D-HST photometric catalog. Furthermore, alternatively using the \SFRuvtot\ derived only from the resolved photometry---rather than all unresolved filters available through the 3D-HST photometric catalog---in the subsequent analysis does not significantly alter the primary conclusions of this paper. Therefore, for the remainder of the analysis we use the more reliable \SFRuvtot\ derived from the 3D-HST photometric catalog, which also has the benefit of being more comparable to other studies.

\subsection{Difference in \EBV\ vs. SFR}\label{sec:EBVdiff}
\begin{figure*}
\begin{adjustbox}{width=\linewidth, center}
\begin{minipage}{\linewidth}
\includegraphics[width=\linewidth]{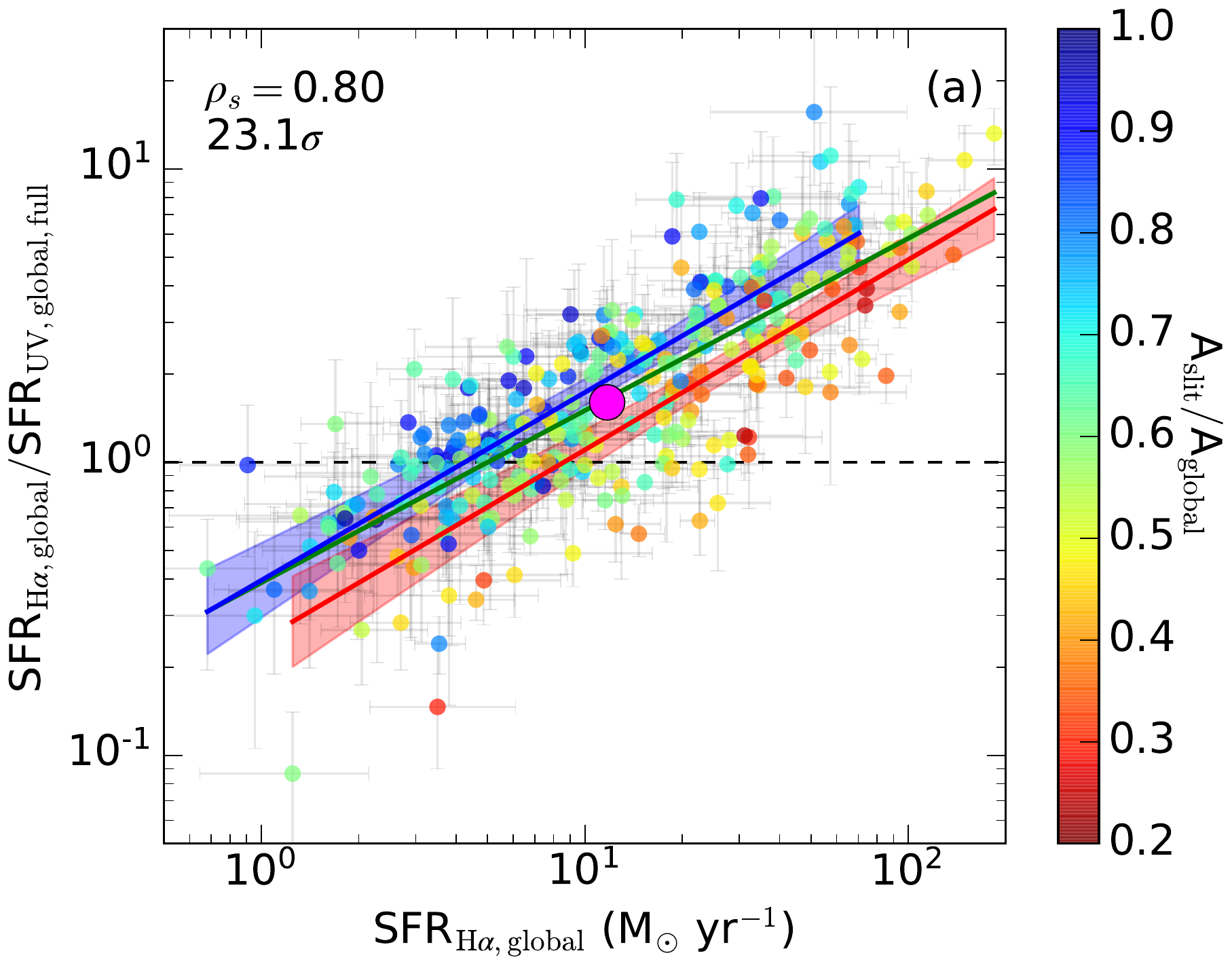}
\end{minipage}
\quad
\begin{minipage}{\linewidth}
\includegraphics[width=\linewidth]{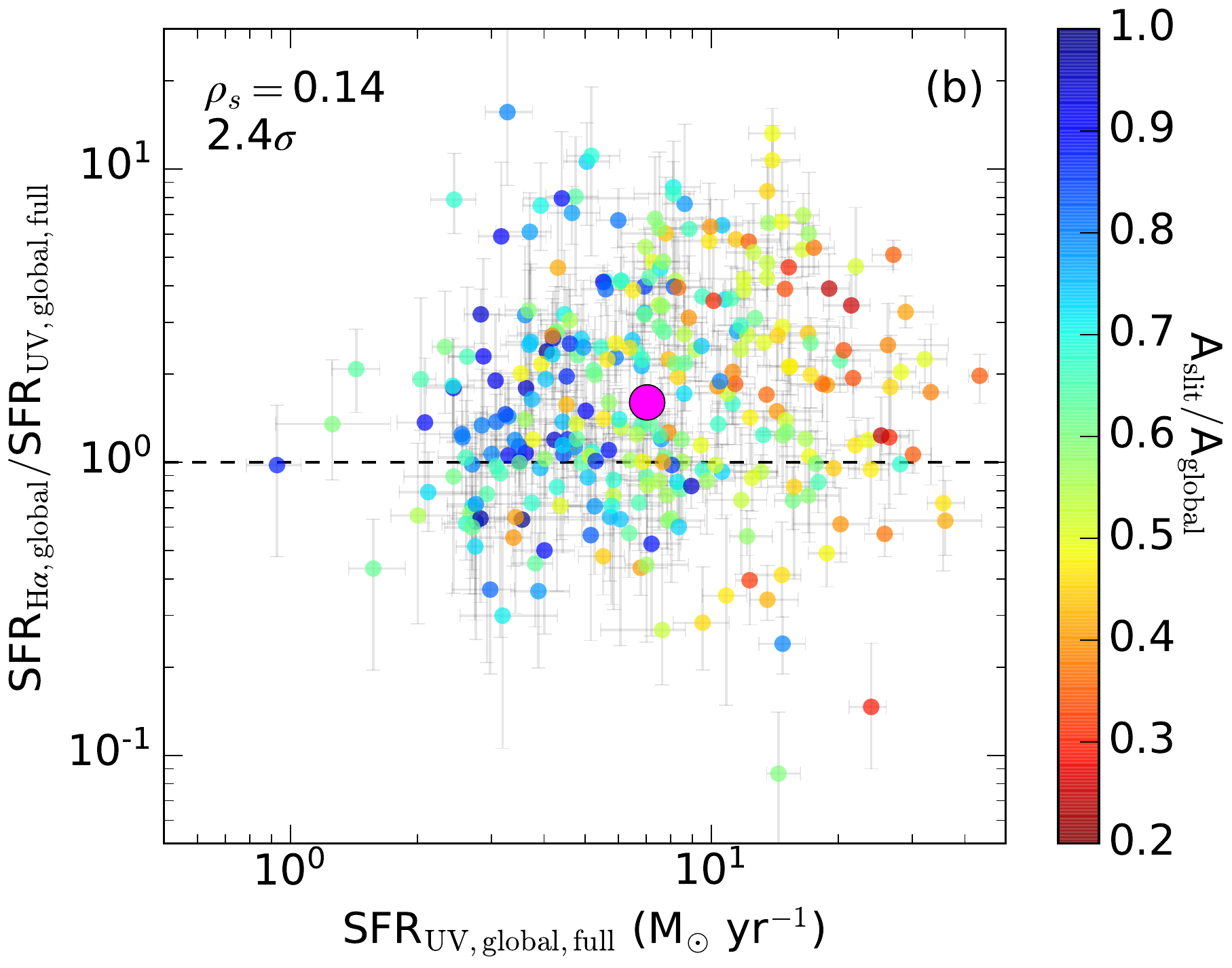}
\end{minipage}
\end{adjustbox}
\begin{adjustbox}{width=\linewidth, center}
\begin{minipage}{\linewidth}
\includegraphics[width=\linewidth]{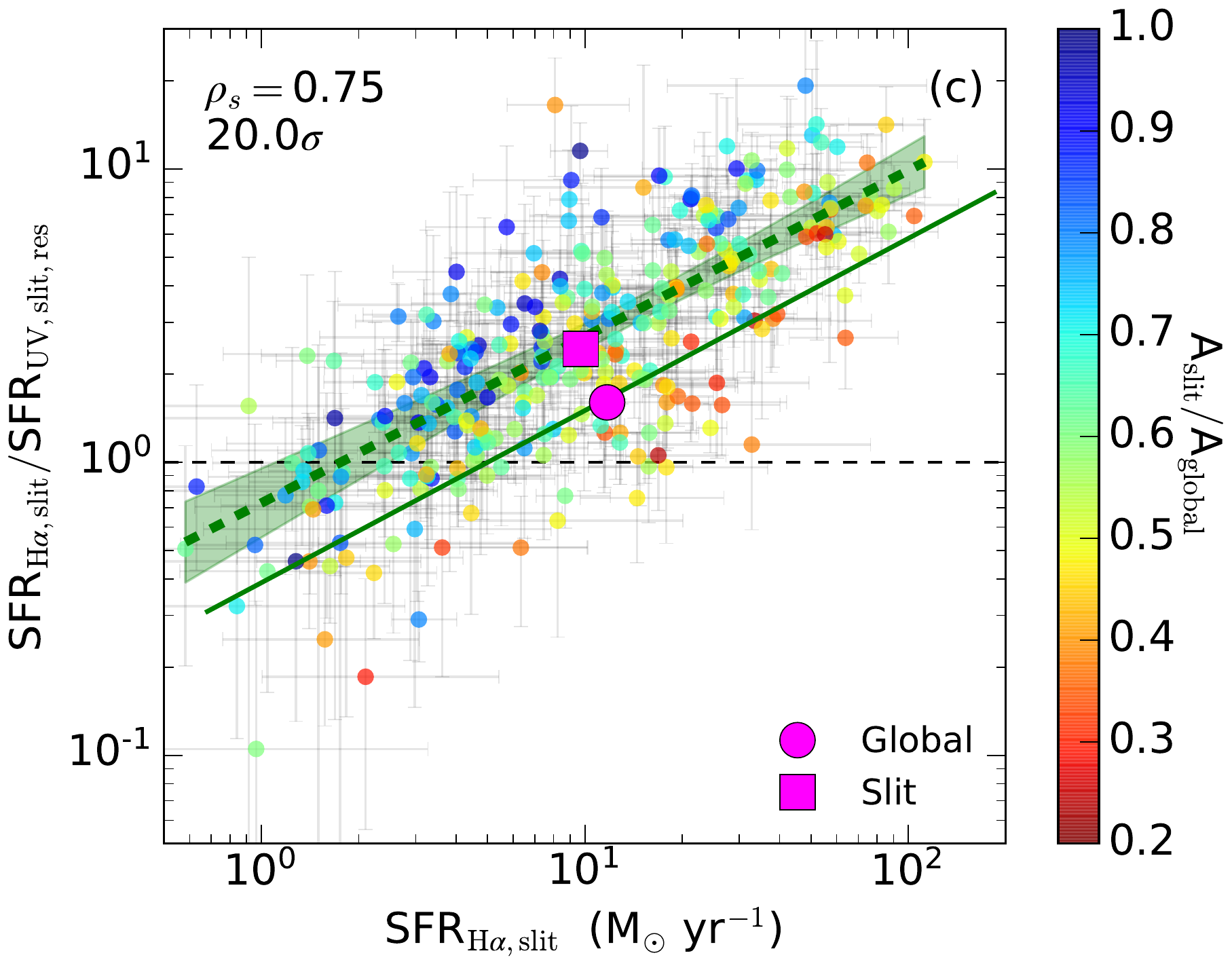}
\end{minipage}
\quad
\begin{minipage}{\linewidth}
\includegraphics[width=\linewidth]{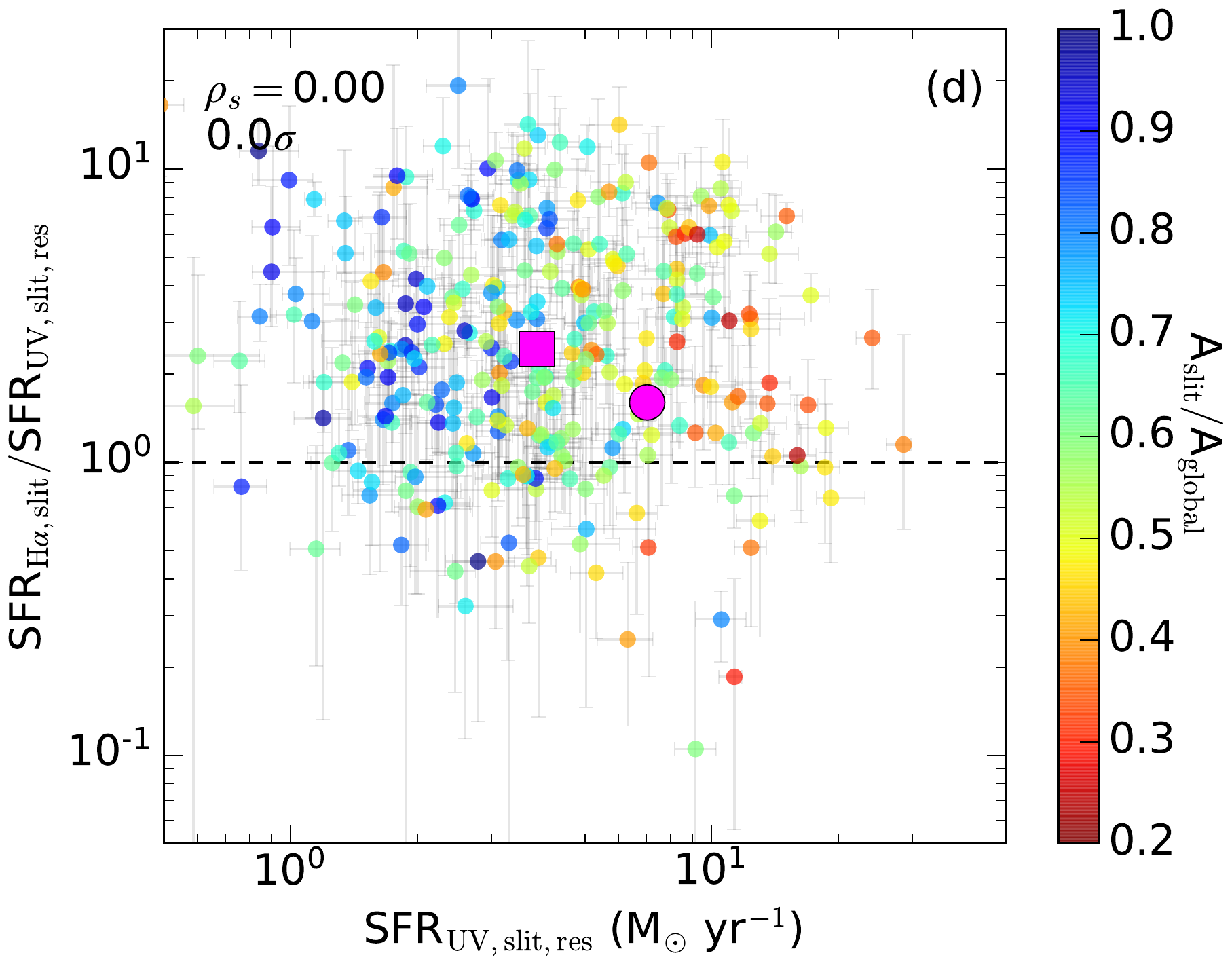}
\end{minipage}
\end{adjustbox}
\caption{\textit{Top:} Globally measured dust-corrected \HA-to-UV SFR ratios versus dust-corrected \SFRha\ (\textit{left}) and \SFRuv\ (\textit{right}) for the \nsamp\ galaxies in the sample. The black horizontal dashed line indicates where the \SFRha\ would equal \SFRuv\ and the points are colored by the fraction of the galaxy that falls within the slit area (see \autoref{fig:picture}). The solid lines show the best-fit relationships for the full sample (green), compact galaxies with $\slitfrac>0.6$ (blue), and large galaxies with $\slitfrac<0.6$ (red). The blue and red shaded regions show the 3$\sigma$ confidence intervals of the best-fit relationships of the compact and large galaxies, respectively, thus emphasizing how the offset from the relationship between \HA-to-UV SFR ratios and \SFRha\ is dependent on galaxy size. \textit{Bottom:} Same as the top panels, except that all of the dust-corrected SFRs (and their respective inferred reddening) are measured exclusively within the spectroscopic slit area. The purple symbols show the median \HA-to-UV SFR ratio, \SFRha, and \SFRuv\ for when measurements are made globally (circles) or directly within the slit area (squares). The dashed green lines shows the best-fit relationship to the full sample for the slit-measured SFRs and, for reference, the solid green lines show the best-fit relationship to the global measurements of the full sample from the top panels. The green shaded regions show the 3$\sigma$ confidence interval of the slit-measured best-fit relationships for all galaxies, thus emphasizing the significance of the shift in normalization between SFR measurements made exclusively within the slit area (bottom panels) compared to those made across the entire galaxy (top panels). The Spearman rank correlation coefficient and its significance are listed in the top left corner of each panel.}
\label{fig:slitfrac}
\end{figure*}
\autoref{fig:EBVdiff_compare} shows how the difference between nebular and stellar continuum reddening varies as a function of SFR, both when all properties are inferred from global flux measurements and when they are inferred from flux measurements restricted to the spectroscopic aperture. The Spearman rank correlation coefficient and the significance of the correlation is listed in the top left corner of each panel. The top panels show the difference between \EBVgtot\ and \EBVstot\ versus \SFRhatot\ (panel a) and \SFRuvtot\ (panel b) when slit-loss corrections are applied to the emission line measurements and the SED is fit to the 3D-HST photometry covering the entire galaxy. The \texttt{scipy.odr} package is used to perform orthogonal distance regression linear fits and obtain 3$\sigma$ confidence intervals on the relationship between differences in reddening and SFR for the top panels of \autoref{fig:EBVdiff_compare} (solid green lines and shaded regions). The observed correlations are partially caused by the SFR being dust-corrected using \EBV. Specifically, the dust corrections cause \SFRha\ to be dependent on \EBVg\ (see \autoref{sec:ebvsfr}) and \SFRuv\ to be dependent on \EBVs. However, the reddening and SFR derived from the nebular emission (\EBVg\ and \SFRha) are independent from those derived from the UV continuum (\EBVs\ and \SFRuv) since they are measured separately from the spectroscopy and photometry, respectively. The bottom panels of \autoref{fig:EBVdiff_compare} show how the difference in \EBVgslit\ and \EBVsslit\ varies with \SFRhaslit\ (panel c) and \SFRuvslit\ (panel d) when slit-loss corrections are not applied and the SED is only fit to the photometric flux with spectroscopic slit coverage (see \autoref{sec:slit_smooth}). The best-fit linear relationships based on the reddening and SFRs derived globally (solid green lines) and exclusively within the slit (dashed green lines) are both shown in the bottom panels in order to demonstrate the changes between the top (global) and bottom (slit) panels. \autoref{fig:EBVdiff_compare} shows that the general relationship between the difference in reddening probes and SFR is statistically significant (23$\sigma$ for \SFRha\ and 5$\sigma$ for \SFRuv) and persists regardless of whether measurements are made directly within the slit or inferred globally, suggesting that aperture effects have little impact on the significance of the correlation (see \autoref{sec:discussion} for an in-depth discussion). In both cases \EBVg\ is on average higher than \EBVs\ (0.16\,mag globally and 0.13\,mag in the slit), with agreement at lower SFRs and a larger discrepancy at higher SFRs. 

\subsection{\HA-to-UV SFR Ratios}\label{sec:slitfrac}
\autoref{fig:slitfrac} shows the relationship between \HA-to-UV SFR ratios versus \SFRha\ (left panels) and \SFRuv\ (right panels) when SFRs are either derived from the entire galaxy area (top panels) or exclusively within the spectroscopic slit area (bottom panels). As can be seen by the median \HA-to-UV SFR ratio (purple symbols), \SFRha\ is typically higher than \SFRuv\ (0.22\,dex for global measurements and 0.39\,dex within the slit). Furthermore, \autoref{fig:slitfrac} shows that \SFRha\ and \SFRuv\ tend to generally agree in galaxies with lower \SFRha\ and deviate more significantly as \SFRha\ increases. The persistence of the correlation between \HA-to-UV SFR ratios and \SFRha\ when all SFRs are measured inside the slit further implies that the trend is not induced by aperture effects. The solid and dashed lines in \autoref{fig:slitfrac} show the best-fit linear relationships between \HA-to-UV SFR ratios and \SFRha\footnote{The best-fit linear relationships are not shown for the correlation between \HA-to-UV SFR ratios and \SFRuv\ since the are not significantly correlated ($<$3$\sigma$).} for the full sample (green), compact galaxies with $\slitfrac>0.6$ (blue), and large galaxies with $\slitfrac<0.6$ (red), where 0.6 is the average area fraction covered by the slit of all galaxies in the sample. To show how the relationship changes between the global and slit measurements, the best-fit relationship to the full sample from \autoref{fig:slitfrac}a is also shown as a solid green line in panel (c). We find that the slope (panel c) and scatter between \HA-to-UV SFR ratios and either probe of SFR are approximately the same regardless of the area over which SFRs are measured (the scatter is 0.22\,dex for global and 0.25\,dex within the slit). The primary difference between global (top panels) and slit-measured (bottom panels) SFRs is the shift between the median points (purple) and, in particular, the shift in the normalization of the trend between \HA-to-UV SFR ratios and \SFRha\ (panel c; dashed green line compared to the solid green line). The shift in normalization is partly due to lower SFR within the slit region, but there is also an additional shift in the \HA-to-UV SFR ratio, which we further discuss in \autoref{sec:discussion}. Alternatively investigating the difference between nebular and stellar continuum reddening and \HA-to-UV SFR ratio versus stellar mass, \EBVg, or \EBVs\ produces similar results due to the well-known SFR--$M_*$ relationship (see \autoref{fig:sample}) and since the dust-corrected SFR is dependent on \EBV\ by definition, but with additional uncertainty for galaxies with low reddening when \EBV\ is consistent with (and assumed to be) zero. 
\begin{figure*}
\includegraphics[width=.7\textwidth]{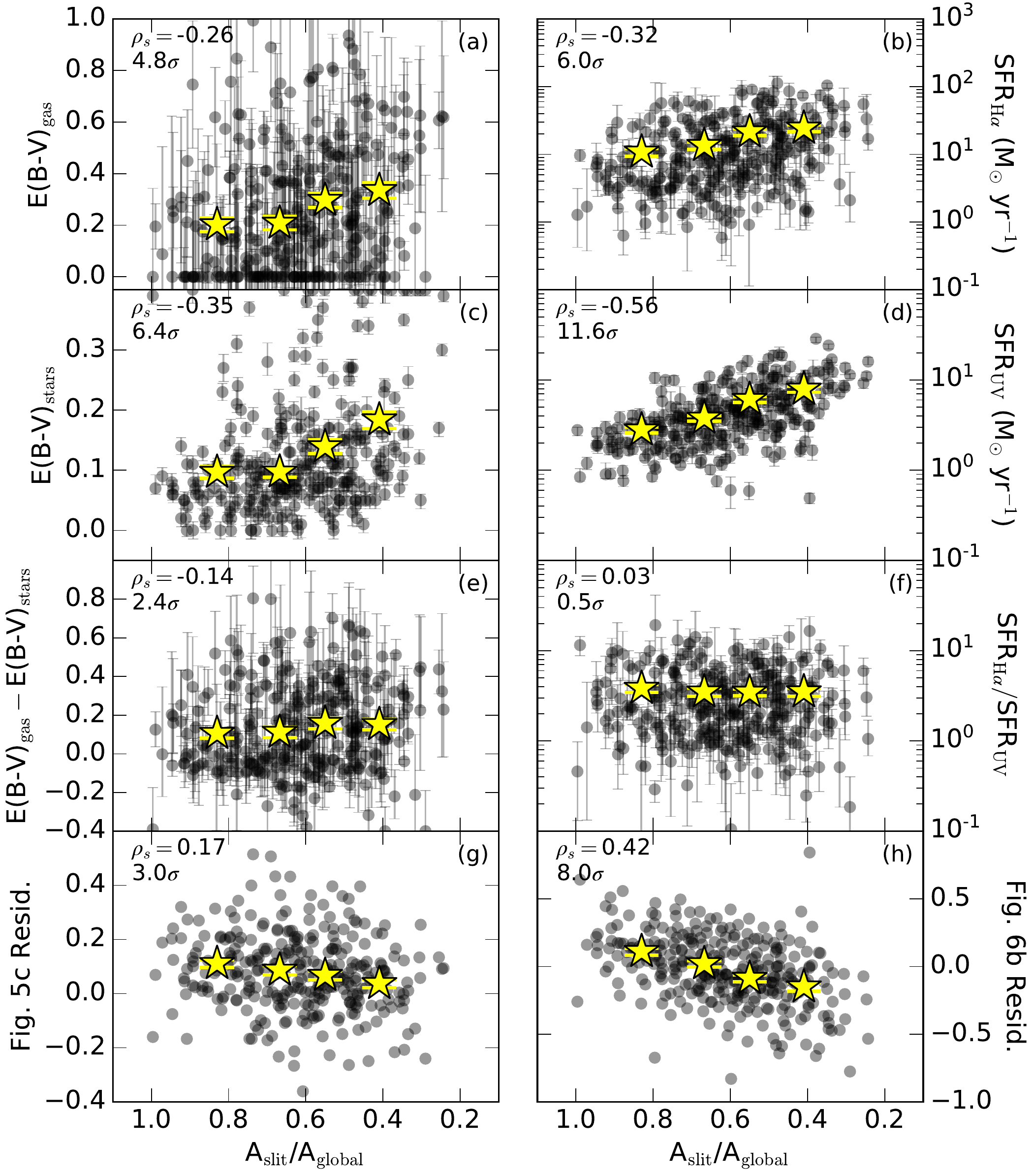}
\caption{Correlations between the fraction of the galaxy within the slit area (i.e., galaxy size) and the following measurements that were all made directly within the slit area: nebular reddening (panel a), \SFRha\ (panel b), stellar continuum reddening (panel c), \SFRuv\ (panel d), the difference between nebular and stellar continuum reddening (panel e), \HA-to-UV SFR ratio (panel f), the residuals from the best-fit linear relation (dashed green line) shown in \autoref{fig:EBVdiff_compare}c (panel g), and the residuals from the best-fit linear relation (dashed green line) shown in \autoref{fig:slitfrac}b (panel h). Since compact galaxies have larger fractional slit areas (\slitfrac\ $\approx1$) than physically large galaxies, the x-axis labels have been reversed. The yellow stars are the averages from individual measurements equally divided into four bins of the fractional slit area (\slitfrac). The Spearman rank correlation coefficient and its significance are listed in the top left corner of each panel.}
\label{fig:slitfrac_trends}
\end{figure*}

\subsection{\EBV\ and SFR vs. Size}\label{sec:areatrends}
In Figures 3 through 6, it can be seen that \slitfrac\ (color-coded points; i.e., galaxy size) is correlated with various plotted quantities. In particular, the best-fit linear relationships shown in \autoref{fig:slitfrac}a indicate that the normalization of the trend between \HA-to-UV SFR ratios and \HA\ SFRs is dependent on size, where the blue and red linear fits represent compact and large galaxies, respectively. In this section we analyze the significance of galaxy size on the relationships discussed in \autoref{sec:EBVdiff} and \autoref{sec:slitfrac} using the Spearman rank correlation test. 

\autoref{fig:slitfrac_trends} shows how \slitfrac\ (i.e., galaxy size) correlates with \EBVgslit\ (panel a), \SFRhaslit\ (panel b), \EBVsslit\ (panel c), \SFRuvslit\ (panel d), the difference between \EBVgslit\ and \EBVsslit\ (panel e), slit-measured \HA-to-UV SFR ratios (panel f), and the residuals from the best-fit linear relationships (dashed green lines) shown in \autoref{fig:EBVdiff_compare}c and \autoref{fig:slitfrac}c (panels g and h). All measurements in \autoref{fig:slitfrac_trends} are made directly within the slit in order to avoid the added complexity to the interpretation when slit-loss correction assumptions are applied. 

We find that light is more dust reddened in larger galaxies (lower \slitfrac; 5$\sigma$ for \EBVg\ and 6$\sigma$ for \EBVs; panels a and c) and that larger galaxies have higher SFRs (6$\sigma$ for \SFRha\ and 12$\sigma$ for \SFRuv; panels b and d) compared to compact galaxies. The two probes of dust reddening also tend to be more discrepant in larger galaxies, with nebular reddening being higher than that of the stellar continuum (2.4$\sigma$; panel e). At a fixed \SFRha, on the other hand, compact galaxies tend to exhibit a more significant difference between nebular and stellar continuum reddening than large galaxies (3$\sigma$; panel g). However, the trend observed in panel (g) may be exaggerated since \SFRha\ depends on \EBVg\ (see \autoref{sec:discussion} for an in-depth discussion). Similarly, when considering that \HA-to-UV SFR ratios are generally insensitive to galaxy size (0.5$\sigma$; panel f) and that \SFRha\ is fixed for a range of \HA-to-UV SFR ratios in \autoref{fig:slitfrac}b, it follows that the trend between galaxy sizes and the residuals between \HA-to-UV SFR ratios and \SFRha\ (8$\sigma$; panel h) is likely caused by the strong correlation between \SFRuv\ and galaxy size (12$\sigma$; panel d). Finally, we note that all of these correlations persist and become more significant when all measurements are alternatively made globally across the galaxy, again suggesting that these relationships are not caused by aperture effects.

\section{Discussion}\label{sec:discussion}
%
%
%
%

\subsection{Aperture Effects}\label{sec:slit_explained}
To interpret the results of our analysis, we first need to assess how the coverage of the spectroscopic slit with respect to the surface area of the galaxies influences the results. We consider \slitfrac\ as a proxy for physical size of the galaxies in our sample since a given angular size corresponds to an approximately fixed physical size over the redshift range covered by our sample ($\zmin<z<\zmax$) and all galaxies are observed using a fixed 0\farcs7 slit width. To remind the reader, \autoref{fig:picture} shows the observed and unobserved emission from a compact and large galaxy relative to the spectroscopic slit boundaries. Galaxies that are nearly entirely contained within the slit boundaries (i.e., the compact galaxies) have a \slitfrac\ near unity, and thus the slit-measured \EBVg\ and \SFRha\ of compact galaxies will be close to those that would be measured globally---if such measurements existed. Large galaxies with lower \slitfrac, on the other hand, require larger corrections for light lost outside of the slit area. The changes between slit-measured and globally inferred quantities for \EBVg\ (\autoref{fig:totslit_compare}a) and \SFRha\ (\autoref{fig:totslit_compare}b) are influenced by the assumptions made when correcting for slit-loss. The usual assumptions when correcting for spectroscopic slit-loss is that the measured Balmer decrement (\HA/\HB) is representative of the average nebular reddening across the galaxy and that the fraction of \HA\ light lost outside the slit is equal to the fraction of $H_{160}$ light lost outside the slit. On the other hand, both the slit and global quantities for \EBVs\ (\autoref{fig:totslit_compare}c) and \SFRuv\ (\autoref{fig:totslit_compare}d) can be directly measured from the resolved CANDELS/3D-HST photometry. 

The top panels of \autoref{fig:EBVdiff_compare} show that \EBVgtot\ and \EBVstot\ become more discrepant as the global SFR increases, and the top panels of \autoref{fig:slitfrac} show that \SFRhatot\ is higher than \SFRuvtot, particularly in galaxies with higher \SFRhatot\ (panel a). These observations could be explained if the globally averaged \EBVg\ or the total \HA\ light is overestimated by the slit-loss corrections in galaxies with high SFRs. If the slit-loss corrections lead to overestimated \EBVg\ and/or \SFRha, then the trends observed in the top panels of \autoref{fig:EBVdiff_compare} and \autoref{fig:slitfrac}a would not be present when the slit-loss corrections are not incorporated into the spectroscopic measurements. However, the slit-measured quantities shown in the bottom panels of \autoref{fig:EBVdiff_compare} and \autoref{fig:slitfrac} show that \textit{there is no significant change in the significance or slope of these relationships}. The \EBVg\ is larger than \EBVs\ at high SFRs, and \SFRha\ remains higher than \SFRuv\ at high \SFRha. Furthermore, at a fixed \SFRha\ we found a systematic shift towards lower differences between \EBVg\ and \EBVs, and lower \HA-to-UV SFR ratios for larger galaxies. Therefore, we conclude that these relationships must reflect the physical properties of the galaxies in our sample.

The best-fit linear relationships included in the bottom panels of \autoref{fig:EBVdiff_compare} and \autoref{fig:slitfrac}c are used to compare the changes between when all quantities are either inferred globally (solid green lines) or measured directly within the slit area (dashed green lines). If \EBVg, \EBVs, and the \HA-to-UV SFR ratio are equivalent within the slit area and globally across the entire galaxy---which is the assumed behavior when estimating the slit-loss corrections that are described in \autoref{sec:slitloss}---then \textit{in all cases these best-fit linear relationships are expected to shift towards lower \HA\ and UV SFR} by definition when the slit-loss corrections are reversed ($\SFRslit<\SFRtot$; solid to dashed green lines; also see right panels of \autoref{fig:totslit_compare}). By comparing the solid and dashed green lines in the bottom panels of \autoref{fig:EBVdiff_compare}, it can be seen that there is no significant change in the slope or normalization of the relationship between the difference in reddening probes versus SFR. By comparing the median of the sample when all measurements are made globally (purple circles) versus exclusively inside the slit (purple squares), it can be seen that there is a shift towards lower \SFRha\ (as expected) and lower $\EBVg-\EBVs$ that causes the two relationships to exhibit a similar slope and normalization. The shift towards lower $\EBVg-\EBVs$ is caused by the 29~galaxies that are $>$2$\sigma$ redder (higher) in \EBVsslit\ than their \EBVstot\ (\autoref{fig:totslit_compare}c), which essentially cancels the shift towards lower SFRs. \autoref{fig:slitfrac}c shows no significant change in the slope of the relationship between \HA-to-UV SFR ratios and \SFRha, but there is a significant shift in the normalization of the best-fit linear relationship towards lower SFRs and higher \HA-to-UV SFR ratios when all measurements are made within the slit compared to those measured globally. In this case, in addition to the leftwards shift caused by \SFRhaslit\ being less than \SFRhatot, the median points also show a significant shift towards higher \HA-to-UV SFR ratios. This significant normalization shift towards higher \HA-to-UV SFR ratios is caused by the relative change between global and slit-measured SFRs not being equivalent between \SFRha\ and \SFRuv\ (0.10\,dex and 0.27\,dex, respectively; see right panels of \autoref{fig:totslit_compare}), as is assumed for constant \HA-to-UV SFR ratios in the slit-loss corrections, and could possibly be indicative of differences between the \HA\ and UV light profiles in these galaxies (see \autoref{sec:SFRtrends} for further discussion). The observed normalization shift in \autoref{fig:slitfrac}c is less significant when \SFRuvtot\ is alternatively derived exclusively from the resolved CANDELS/3D-HST photometry (\autoref{fig:bbintSFR}) since these measurements are typically only 0.10\,dex higher than \SFRuvslit, but the observed normalization offset towards higher \HA-to-UV SFR ratios is still significant by $>$3$\sigma$ for $\SFRha \gtrsim 2$\,\Msun\,yr$^{-1}$ (see right panel of \autoref{fig:bbintSFR}).

\subsection{Differences between \EBVg\ and \EBVs}\label{sec:EBVtrends}
The bottom panels of \autoref{fig:EBVdiff_compare} show that the observed differences between \EBVg\ and \EBVs\ persist in galaxies with higher \SFRha\ (panel c) and \SFRuv\ (panel d), even when all measurements are made exclusively within the spectroscopic slit area. The increasing difference between nebular and stellar continuum \EBV\ with SFR may imply that the dust distribution is patchier in galaxies with higher SFRs \citep[which are also inherently dustier;][]{Calzetti94, Boquien09, Price14, Reddy15}, such that the nebular emission becomes dominated by the most obscured massive stars. However, \SFRha\ and \SFRuv\ are not independent from \EBVg\ or \EBVs, in that they are each used to correct these SFR indicators for the effects of dust. Therefore, the observed trends between the \EBV\ difference and SFR in \autoref{fig:EBVdiff_compare}, and similarly the trends observed in panels (e) and (g) of \autoref{fig:slitfrac_trends} (due to the significant correlation between \slitfrac\ and \SFRuv\ shown in panel d), may be exaggerated rather than being caused by the intrinsic characteristics of these galaxies. Furthermore, \citet{Shivaei20} found that the relationship between the differences in \EBV\ versus SFR (\HA\ and UV) for $z\sim2$ star-forming galaxies in the MOSDEF sample is less significant when the UV light is corrected using a metallicity-dependent dust attenuation curve. Therefore, understanding the nature of the difference between nebular and stellar continuum reddening would be best addressed using a SFR indicator that is independent from \EBVg\ and \EBVs, such as total SFRs computed by combining unobscured UV and IR continuum measurements.

\subsection{Variations in \HA-to-UV SFR Ratios}\label{sec:SFRtrends}
\autoref{fig:slitfrac}c shows that the relationship between \HA-to-UV SFR ratios and \SFRha\ persists when all reddening and dust-corrected SFRs measurements are made directly inside the spectroscopic slit area, implying that the trend between \HA-to-UV SFR ratios and \SFRha\ is not driven by the slit-loss corrections. Furthermore, the offset of individual galaxies from the best-fit linear relation between \HA-to-UV SFR ratios and \SFRha\ (green lines) is significantly correlated with the fraction of the galaxy that is directly observed within the spectroscopic slit area (i.e., galaxy size; see \autoref{fig:picture} and \autoref{fig:slitfrac_trends}h). By separately identifying the significance of the correlations between galaxy size and slit-measured \SFRha, \SFRuv, \HA-to-UV SFR, and the residuals from the best-fit relationship between \HA-to-UV SFR ratios versus \SFRha\ (right panels of \autoref{fig:slitfrac_trends}), we find that \SFRuv\ is most significantly correlated with galaxy size (panel d) and that \HA-to-UV SFR ratios are generally constant with galaxy size but with large scatter (panel f). It is well-known that larger galaxies tend to have higher SFRs on average compared to compact galaxies \citep[panels b and d of \autoref{fig:slitfrac_trends};][]{Toft07, Trujillo07, van_Der_Wel14, Scott17, van_de_Sande18}. However, further taking into account that the correlation between galaxy size and SFR is stronger in the UV than \HA, then it is expected by definition that, at a fixed size (or \SFRuv) the \HA-to-UV SFR ratio will be higher for galaxies with higher \SFRha\ relative to galaxies with lower \SFRha\ at the same size (blue, green, and red lines in \autoref{fig:slitfrac}a). Therefore, we reiterate how understanding the differences between \HA\ and UV SFRs would be best addressed using independently derived galaxy properties, such as SFRs measured from the far-IR continuum. 

Generally, the \HA-to-UV SFR ratio may reflect variations in the SFHs of galaxies \citep[][]{Reddy12-1, Price14, Madau14}. High \HA-to-UV SFR ratios, which we observe in galaxies with SFRs $\gtrsim$2\,\Msun\,yr$^{-1}$, could be indicative of SFRs that change on timescales $<$100\,Myr. However, a single rapid burst, or galaxies with complex SFHs, may not sufficiently affect the \HA-to-UV luminosities in order to explain the observed SFR discrepancies \citep[e.g.,][]{Emami19}. Furthermore, observations of local galaxies have found that bursty star formation is more typical of low-mass dwarf galaxies \citep[$<$10$^9$\,\Msun;][]{Weisz12, Guo16, Emami19}, which are lower in mass than the majority of the galaxies in our sample. We find better agreement between \SFRha\ and \SFRuv\ for galaxies with lower SFRs, which corresponds with the lowest mass galaxies in our sample (\autoref{fig:sample}). \autoref{fig:slitfrac_trends}f suggests that \HA-to-UV SFR ratios measured from compact galaxies are consistent with those measured from the centers of large galaxies, such that \HA-to-UV SFR ratios may not be correlated with the physical sizes of galaxies. On the other hand, \HA\ and UV light profiles may not necessarily follow each other in the outskirts of large galaxies, which are beyond the capability of the MOSDEF observations used in the current study (without making assumptions for spectroscopic slit-loss). Centrally peaked \HA\ and Balmer decrement radial profiles have been observed in galaxies at $1<z<3$ using resolved emission line maps \citep[e.g.,][]{Nelson13, Hemmati15, Tacchella18}. Furthermore, massive star-forming galaxies, which are well-known to be larger \citep[e.g.,][]{Toft07, Trujillo07, van_Der_Wel14, Tacchella15, Scott17, Isobe20}, dustier \citep[e.g.,][]{Reddy06, Reddy10, Pannella09, Price14, Shivaei20}, and have higher SFRs \citep[e.g.,][]{Noeske07, Daddi07, Pannella09, Wuyts11, Reddy12-1, Whitaker12, Whitaker14, Shivaei15} than low-mass galaxies on average, exhibit stronger centrally peaked \HA\ and Balmer decrement radial profiles than low-mass galaxies \citep{Nelson16}. If the \HA\ emission is centrally peaked in a galaxy relative to its $H_{160}$ light profile, then the typical assumptions when correcting for slit loss may result in overestimated \SFRha. In the MOSDEF survey, we can only observe a small fraction of the surface area of large galaxies spectroscopically (see \autoref{fig:picture}). However, it is possible that these large galaxies have higher \SFRha\ in their centers compared to their unobserved outskirts (normalization shift between the green lines in \autoref{fig:slitfrac}c), which could be indicative of inside-out galaxy growth \citep[e.g.,][]{Patel13, Gomes16, Nelson16, Jafariyazani19}. Similarly, if Balmer decrements are typically higher in the centers of large galaxies compared to their outskirts, then the assumed average reddening based on the slit measurements will overestimate the globally-averaged reddening. 

Deviations in the attenuation curve relative to the one assumed can cause large systematic uncertainties such that derivations of certain stellar population properties may require more realistic attenuation curves for high-redshift galaxies, such as a mass or metallicity dependent attenuation curve \citep[e.g.,][]{Reddy06, Reddy10, Reddy12, Reddy18, Noll09, Siana09, Kriek13, Shivaei18, Shivaei20}. \citet{Reddy18} found that alternatively assuming a \citet{Calzetti00} attenuation curve over an SMC extinction curve \citep{Fitzpatrick90, Gordon03} causes a systematic shift in absolute masses---but not their relative order---and that assuming an SMC attenuation curve and sub-Solar metallicities is generally more appropriate for young, high-redshift galaxies, such as those in our sample (see \autoref{sec:sedfit}). We consider variations in the dust-to-star geometry relative to that which is already assumed by the choice of attenuation curve that could explain the observed discrepancies between inferred \EBV\ and dust-corrected SFRs. The increase in \HA-to-UV SFR ratios with SFR, for example, may be a result of a patchier distribution of dust in galaxies with higher SFRs. In this scenario, stars of all masses are obscured by roughly the same columns of dust at low SFR, while the younger stellar populations will generally be more obscured in higher SFR galaxies \citep[e.g.,][]{Reddy15}. However, a patchy dust distribution could also cause some sightlines towards older OB associations that contribute to the nebular emission to be less obscured than those still enshrouded by their birth clouds. Galaxies with patchier dust distributions (or centrally peaked \SFRha\ and/or Balmer decrements) may also exhibit \HA\ and UV light profiles that do not match at all radii in large galaxies. Inappropriate assumptions about the shape of the \HA\ and UV light profiles relative to each other may explain: 1) the normalization shift between \HA-to-UV SFR ratios that is observed inside the slit compared to global measurements (\autoref{fig:slitfrac}c), and 2) the correlation between the size of a galaxy and the scatter in the relationship between \HA-to-UV SFR ratios and \SFRha\ (\autoref{fig:slitfrac_trends}h). We will further explore the distribution of dust and its relationship to globally measured properties of the $z\sim2$ MOSDEF galaxies in a future work by performing an in-depth analysis of the reddening maps created through the procedures described in \citet{Fetherolf20}.

\section{Summary}\label{sec:summary}
In this study, we used a sample of \nsamp\ star-forming galaxies drawn from the MOSDEF survey at spectroscopic redshifts $\zmin<z<\zmax$ in order to directly compare nebular and stellar continuum dust reddening, and dust-corrected SFRs measured from \HA\ emission and the UV continuum light. We used the CANDELS/3D-HST high-resolution, multi-band imaging to measure the stellar population and dust properties both globally and within the MOSFIRE slit region. By combining the slit-measured stellar population and dust properties from the resolved imaging with the MOSDEF \HA\ and \HB\ emission line measurements, we were able to directly obtain measurements of \EBVg, \EBVs, \SFRha, and \SFRuv\ exclusively inside the MOSFIRE spectroscopic slit area. 

In order to directly compare photometric and spectroscopic measurements over the same area, the \HST\ imaging was PSF-smoothed to match the spatial resolution (i.e., seeing) of the MOSFIRE spectroscopic observations (\autoref{sec:slit_smooth}). By directly comparing \EBV\ and SFR measurements within the spectroscopic slit area, we found that \textit{aperture corrections alone cannot explain the observed differences between \HA\ and UV probes of reddening and SFR} that are most discrepant in galaxies with high SFR (\autoref{fig:EBVdiff_compare} and \autoref{fig:slitfrac}). We also found that the offset from the best-fit linear relationship between \HA-to-UV SFR ratios and \SFRha\ is significantly correlated with galaxy size (\autoref{fig:slitfrac}a and \autoref{fig:slitfrac_trends}h), but could be attributed to a stronger correlation between galaxy size and \SFRuv\ compared to \SFRha\ (panels b and d of \autoref{fig:slitfrac_trends}). Furthermore, the shift in normalization in the relationship between \HA-to-UV SFR ratios and \SFRha\ when all measurements are made directly within the slit compared to those made globally (\autoref{fig:slitfrac}c) suggests that \SFRha\ may be higher in the centers of large galaxies compared to their outskirts. Assuming that the slit-loss corrections are adequate, higher \SFRha\ in the centers of large galaxies (where there is spectroscopic slit coverage) compared to their outskirts could be suggestive of inside-out galaxy growth. 

To explain the observations presented in \autoref{sec:slit}, we reiterate the physical scenario depicted by \citet{Reddy15} where the dust distribution gets patchier as a function of increasing SFR. In this scenario the optical depth is on average higher in high-SFR galaxies, but the emitted light is dominated by young, massive stellar populations that may be less obscured than what is implied by the average measured reddening---thus leading to the discrepancy between \HA\ and UV SFRs. Furthermore, large galaxies with higher SFRs may exhibit centrally peaked \HA\ or Balmer decrement profiles \citep[e.g.,][]{Nelson13, Hemmati15, Tacchella18} such that their relative \HA\ and UV light profiles are not necessarily equivalent at all radii. 

Due to the nature of the MOSFIRE spectrograph, we can only draw direct conclusions based on the observations that fall within the slit area. Even with our limited spectroscopic capabilities, we see evidence that the observed discrepancies in \EBV\ and SFRs measured from the nebular emission and the stellar continuum are caused by some physical driver---such as centrally peaked or patchy dust distributions---rather than systematic errors in the aperture corrections. These results can be improved upon with independent confirmation of the slit-loss corrections through IFU observations, high spatial resolution spectroscopy that could be used to compare \HA\ and UV emission on smaller scales and out to larger radii, and pairing this analysis with a third independent SFR indicator (e.g., measured from the far-IR continuum). In direct follow-up to this work, we will investigate how the distribution of dust influences the observed SFRs and other globally measured physical properties of high-redshift galaxies.

\section*{Acknowledgements}
This work is based on observations taken by the CANDELS Multi-Cycle Treasury Program and the 3D-HST Treasury Program (GO 12177 and 12328) with the NASA/ESA \HST, which is operated by the Association of Universities for Research in Astronomy, Inc., under NASA contract NAS5-26555. The MOSDEF team acknowledges support from an NSF AAG collaborative grant (AST-1312780, 1312547, 1312764, and 1313171) and grant AR-13907 from the Space Telescope Science Institute. The authors wish to recognize and acknowledge the very significant cultural role and reverence that the summit of Maunakea has always had within the indigenous Hawaiian community. We are most fortunate to have the opportunity to conduct observations from this mountain.
\\ \\
\textit{Facilities:} HST (WFC3, ACS), Keck:I (MOSFIRE), Spitzer (IRAC)
\\ \\
\textit{Software:} Astropy \citep{Astropy_collaboration13, Astropy_collaboration18}, Matplotlib \citep{Hunter07}, NumPy \citep{Oliphant07}, SciPy \citep{Oliphant07}, specline \citep{Shivaei18}

\section*{Data Availability}
Resolved CANDELS/3D-HST photometry is available at \url{https://3dhst.research.yale.edu/Data.php}. Spectroscopic redshifts, 1D spectra, and 2D spectra from the MOSDEF survey are available at \url{http://mosdef.astro.berkeley.edu/for-scientists/data-releases/}. 

\bibliographystyle{mnras}
\bibliography{mapsbib}

\bsp
\label{lastpage}
\end{document}